\documentclass[12pt]{article}
\input epsf.sty
\newcommand{\HH}{{\cal H}}

\newcommand{\nl}{\hspace{-.65cm}}
\newcommand{\be}{\begin{equation}}
\newcommand{\ee}{\end{equation}}
\newcommand{\ben}{\begin{eqnarray}\displaystyle}
\newcommand{\een}{\end{eqnarray}}
\newcommand{\refb}[1]{(\ref{#1})}

\newcommand{\sectiono}[1]{\section{#1}\setcounter{equation}{0}}

\def\sqr#1#2{{\vcenter{\vbox{\hrule height.#2pt
         \hbox{\vrule width.#2pt height#1pt \kern#1pt
            \vrule width.#2pt}
         \hrule height.#2pt}}}}

\begin{document}

{}~ \hfill\vbox{\hbox{hep-th/0304192} \hbox{PUPT-2082} }\break

\vskip 1cm

\begin{center}
\Large{\bf Closed Strings as Imaginary D-branes }

\vspace{10mm}

\normalsize{Davide Gaiotto, Nissan Itzhaki and Leonardo Rastelli}

\vspace{10mm}

\normalsize{\em Physics Department, Princeton University,
Princeton, NJ 08544}\end{center}\vspace{10mm}

\begin{abstract}

\medskip

Sen has recently drawn attention to an exact  time-dependent
Boundary Conformal Field Theory with the space-time interpretation
of  brane creation and annihilation. An interesting limit of this
BCFT  is formally equivalent to an array of D-branes located in
{\it imaginary} time.
 This raises the question: what is the
meaning of D-branes in imaginary time? The answer we propose is
that D-branes in imaginary time define {\it purely closed} string
backgrounds.
 In particular we prove that the
disk scattering amplitude of $m$ closed strings off  an arbitrary
configuration of imaginary branes is {\it equivalent} to a {\it
sphere}  amplitude with $m+1$ closed string insertions. The extra
puncture is a specific closed string state, generically
normalizable, that depends on the details of the brane
configuration. We study in some detail the special case of the
array of imaginary D-branes related to Sen's BCFT and comment on
its space-time interpretation. We point out that a  certain limit
of our set-up allows to study classical black hole creation and
suggests a relation between  Choptuik's critical behavior and a
phase-transition  {\it \` a la} Gregory-Laflamme. We speculate that
open string field theory on imaginary D-branes is dual to string
theory on the corresponding closed string background.

\end{abstract}

\newpage

\baselineskip=18pt

\tableofcontents

\newpage

\sectiono{Introduction }
\bigskip

Worldsheet duality between  open and closed strings
 is one of the  truly fundamental ideas
of string theory. Many modern developments have originated from
the application of this idea to supersymmetric configurations of
D-branes.  In this paper we show that the study of
non-supersymmetric, unstable D-branes gives some clues for a new
intriguing incarnation of open/closed string duality.

\smallskip

Our starting point is the  real-time process of D-brane creation
and annihilation, which has recently attracted much attention
\cite{gut}-\cite{e}. In particular, Sen \cite{cosh} introduced a
simple class of models in bosonic string theory obtained by
perturbing the flat-space $c=26$ CFT with the exactly marginal
deformation
\be \lambda \, \int \,
dt \, \label{iu} \cosh ( X^0(t)) \, , \ee
where $t$ is a coordinate on the worldsheet boundary, and
$\lambda$ is a free parameter in the range $0 \leq \lambda \leq
\frac12$. This is a family of exact solutions of { classical} open
string theory whose space-time picture is that of an unstable
brane being created at a time $X^0 \sim - \tau$ and decaying at a
time $X^0 \sim + \tau$, with $\tau = -\log ( \sin ( \pi
\lambda))$. For $\lambda=\frac12$ the lifetime of the brane is
zero,  that is, there is no brane to be found anywhere. Moreover,
the corresponding boundary state appears to vanish identically
\cite{cosh}. This  is fascinating as it seems to suggest that for
$\lambda =\frac12$ the BCFT \refb{iu} describes the stable closed
string vacuum, where open string degrees of freedom are absent.
Somehow the boundary perturbation \refb{iu} with $\lambda = \frac
12$ must get rid of the worldsheet boundary!

\smallskip

In the framework of
\refb{iu}, we have the opportunity
to precisely test a scenario  \cite{shata, gai} for
how purely closed
string amplitudes may be obtained at the tachyon vacuum.
The basic
idea is that as the tachyon condenses, worldsheets with large
holes are suppressed, and the integration over  moduli space
should localize to the region where the holes shrink to points.
This heuristic picture was made somewhat more concrete in
\cite{gai},
where it was argued that amplitudes for $m$ external closed
strings on the disk reduce at the tachyon vacuum to {\it sphere}
amplitudes with the same $m$ closed string punctures {\it plus} an
additional insertion of a zero-momentum state (possibly a soft
dilaton), a remnant of the shrunk boundary. This analysis \cite{gai}
was performed in the framework of  a regulated version
of vacuum string field theory \cite{vsft}. However, due to subtleties in the
regulation procedure, it was difficult to make
this conclusion completely precise.

\smallskip

Since we wish to focus on the case $\lambda =\frac12$, it is very
useful to realize that at this critical value the BCFT admits a
simple description. By Wick rotation $X^0 \to i X$, one obtains
the well-known exactly marginal deformation $\lambda \cos(X)$
\cite{cal} (infact this is  how Sen arrived at \refb{iu} in the
first place), which for $\lambda=\frac12$ is equivalent to an
infinite array of D-branes located at $X=2\pi(n+\frac12)$
\cite{cal, pol}. We could thus say that at the critical value $\lambda
= \frac12$, the time-dependent boundary deformation \refb{iu}
becomes an array of D-branes located at {\it imaginary} times \be
\label{Array}
 X^0=  i \, 2\pi  (n+\frac12 ) \,  , \quad n \in {\bf Z} \,.
\ee
This is an empty statement if we do not define the meaning of
D-branes  in imaginary time. Our approach to that issue is very
simple: Any quantity that one wishes to compute for a
configuration of D-branes in imaginary time should be obtained by
Wick rotation of the configuration in real space. This
prescription gives consistent
 answers with an interesting physical meaning.

\smallskip

A natural class of observables is given by scattering amplitudes
of closed strings  on the disk in the background of the  brane
configuration \refb{Array}. We find that these reduce to sphere
amplitudes with $m+1$ punctures. The extra puncture is, however,
 {\it not} a soft dilaton as in \cite{gai}, 
but a non-trivial closed string state
that involves the whole tower of massive modes. Thus  the BCFT
\refb{iu} with $\lambda = \frac 12$
 describes a  purely closed string background
with no physical open string degrees of freedom. This background,
however, is not  the closed string  {\it vacuum},
but a  specific time-dependent state with non-zero energy.
The detailed features of this background are  very reminiscent of
`tachyon matter' \cite{s,4}.

\smallskip

While these results were first obtained in the
 special case \refb{Array}, the basic conclusion
 is much more general.
A generic configuration of imaginary D-branes defines   a closed
string background. The details of the background depend on the
details of the configuration of imaginary D-branes. Exactly
marginal open string deformations, for example deformations that
move the positions of the imaginary branes, are naturally
reinterpreted as deformations of the closed string background. The
case \refb{Array} is seen to be very special as the closed string
state has divergent norm \cite{mald}, and the background admits an
additional exactly marginal deformation which is associated with
 the creation of an actual brane in real time.

\smallskip

An outline of the paper is as follows. In section 2 we spell out
the basic prescription for how to deal with the array of D-branes in
imaginary time. We consider a more general case in which $X^0=i
a(n+\frac12 ) $, where $a$ is an arbitrary parameter,
 and find a general formula for
disk  amplitudes associated with such an
array.  In section 3 we analyze in detail  disk amplitudes
for scattering of $m$ external closed strings from the
array of D-branes in imaginary time. By an exact computation, we
show that they are
{\em equivalent} to {\em sphere} amplitudes with the same $m$
closed strings insertions plus the insertion of an extra closed string
state $| W  \rangle$.
The details of this closed string state, in particular
its space-time interpretation,
 are studied  in section 4.  The energy of the state is finite
and of order $O(g_s^0) =O(1)$ for any $a > 2 \pi$.
 The case $a = 2 \pi$,
corresponding to $\lambda = \frac12$ in the BCFT \refb{iu}, is
seen to be special as the normalization and energy of the state
diverge \cite{mald}. We suggest a heuristic mechanism to cutoff
this divergence,
namely we point out that the gravitational
back-reaction makes the effective distance $a_{eff} = 2 \pi +
\gamma \, g_s$, with $ \gamma >0$, leading to a total  energy of
order  $1/g_s$.

\smallskip

In section 5 it is shown that  although the imaginary array of
branes does not have propagating open string degrees of freedom, open
strings still play an important role. There still exists a
discrete set of on-shell open string vertex operators
corresponding to exactly marginal deformations, for example
deformations that move the branes around in imaginary time.  These
open string moduli are re-interpreted as closed string
deformations, according to a precise dictionary. In section 6 we
show that, subject to certain reality conditions, one can
distribute D-branes quite freely in the complex $X^0$ plane. The
reduction of disk amplitudes to sphere amplitudes still holds in
this general case. We briefly comment on  extensions  to
the superstring. In section 7 we briefly discuss some ideas
about the open string field theory associated with D-branes
in imaginary time
and speculate that a version of vacuum string field theory may be
obtained in the limit $a \to \infty$.

\smallskip

 Section 8 is devoted to the case $a=2\pi$. In this case there is
an additional exactly marginal open string deformation
(which we may label as `$\cosh(X^0)$')
 which is {\it not} dual to  a purely closed deformation, as it
introduces instead an actual  D-brane  in real time. We give a
treatment of \refb{iu} for all $\lambda \leq \frac12$ by representing
the boundary state as an infinite array of  some specific {\it
smeared} sources, and then performing the Wick rotation. We find
that for $\lambda < \frac12$ a time-delay of order $\tau$ is
introduced between the incoming and the outgoing parts of the
closed string wave, while for $|X^0|  < \tau$ there is an actual
source for the closed string fields.

\smallskip

Finally in section 9 we describe a curious application of our set-up.
We consider an array
of branes in imaginary time,
and tune various parameters in such a way that
the closed string state $|W \rangle$  becomes simply a {\it classical},
spherically symmetric dilaton wave with barely  enough
energy to form a black hole. Fascinating critical behavior
 was discovered by Choptuik \cite{cho} in such a system. We observe that in the
corresponding Euclidean theory  this critical point corresponds
to a  phase transition reminiscent of the
Gregory-Laflamme \cite{Greg} phase transition. We speculate on a possible
realization of this phase transition in large $N$ open string field theory.

\medskip

While this paper was in preparation we received \cite{mald} that
has some overlapping results.

\sectiono{Preliminaries}

We wish to give a meaning to the notion of an array of D-branes
located at imaginary times $X^0 = i (n + \frac 12) a$. The distance $ a
= 2 \pi$ (in units $\alpha' =1$) corresponds to the $\lambda
=\frac12$ case of the BCFT \refb{iu}, but it is interesting and
not more difficult to keep $a$ arbitrary.  In this section we
define our basic prescription to compute disk amplitudes of
external closed strings in the background of this `imaginary
array'.
 We first give a naive argument
why all such scattering amplitudes vanish.
Then we illustrate, via a simple
example, a natural analytic continuation prescription
that actually yields non-zero answers.
Finally  we derive a simple general formula that expresses
the scattering amplitude $S$ in the background of the imaginary array
 in terms of the scattering amplitude $\widetilde A$ for
a single  D-brane in real time.

\subsection{Naive argument}

Let us denote the scattering amplitude of some closed strings off
a single D-brane located at   $X=0$ by $\widetilde A$. Here $X$ is a spatial
coordinate.  To find the scattering amplitude $S$ for an array of D-branes located
at imaginary time we could proceed in two steps:

\smallskip

\nl (i) Find the scattering amplitude $\widetilde S$ for an array of
D-branes located at the real positions $X=(n + \frac 12) a$, $n \in {\bf Z}$.
This is given by \be\label{i1}
\widetilde S (P, \dots) = \sum_{n=-\infty}^\infty
  {\widetilde A} (P, \dots) e^{i \left(n + \frac{1}{2}\right)a P }
  =\widetilde{A}(P, \dots)
   \sum_{n=-\infty}^\infty (-1)^n  \, 2 \pi \delta(P a-2\pi n) ,\ee
where $P$ is the total momentum in the $X$ direction
and the dots denote other variables that the amplitude may depend on.
 This is a precise equality in the sense of distributions.
Simply put, the momentum  has to be quantized due to the periodicity in $X$.

\smallskip

\nl (ii) Apply (inverse) Wick rotation \be
\label{Wick}
X \rightarrow -i
X^0,\;\;\;P\rightarrow iE,\ee where now $X^0$ and $E$ are real.
This has the effect of turning the spatial coordinate $X$ into
a temporal coordinate $X^0$  and  of rotating the array of D-branes from the
real axis to the imaginary axis, $X^0 = i (n + \frac 12) a$. So it
takes us exactly to the set-up that we wish to study. Wick
rotating (\ref{i1}) we get the formal expression
\be \label{Snaive}
S(E, \dots)\equiv \widetilde S ( i E, \dots) = \widetilde{A}(i E, \dots)
\sum_{n=-\infty}^\infty (-1)^n  \, 2 \pi \delta(iEa-n) .\ee Naively this
implies that $S(E, \dots)$ is identically zero since for any real
$E$ the delta functions vanish. However  one has to be more careful, as
$\widetilde{A}(i E, \dots)$ may blow up for
some real values of $E$ yielding a non-zero
 $S(E)$.  Clearly the discussion so far has been quite formal,
for example the  summation \refb{i1}
does not commute with the Wick rotation \refb{Wick} in the sense that if we
first Wick rotate and then sum over the array,  the sum
does not converge for any  real $E$.  We need to specify an unambiguous
prescription for the analytic continuation \refb{Wick}.
Let us illustrate how  a natural prescription comes about in a simple example.

\subsection{Example}

The example we wish to study is \be \label{i3} \widetilde{A}(P,
\dots)=\frac{1}{P^2+c^2}, \ee where  $c$ is a real number  (we
take for definiteness $c > 0$) that can depend on the other
variables but not on $P$.  This is the one-dimensional Euclidean
propagator. In position space, \be \widetilde G(X)= \int dP \,
\widetilde A(P) \, e^{i PX}  =
 \frac{\pi}{c} e^{-c|X|} \, ,
\ee which obeys $ \left( \frac{d^2}{d X^2} - c^2 \right)
\widetilde G(X) = -\delta(X) \,.$
 All the amplitudes we study in this paper can be expanded in
terms of (\ref{i3}). Hence this example is of special importance.
\begin{figure}
\begin{picture}(240,232.5)(0,0)
\vspace{-5mm} \hspace{10mm} \mbox{\epsfxsize=130mm
\epsfbox{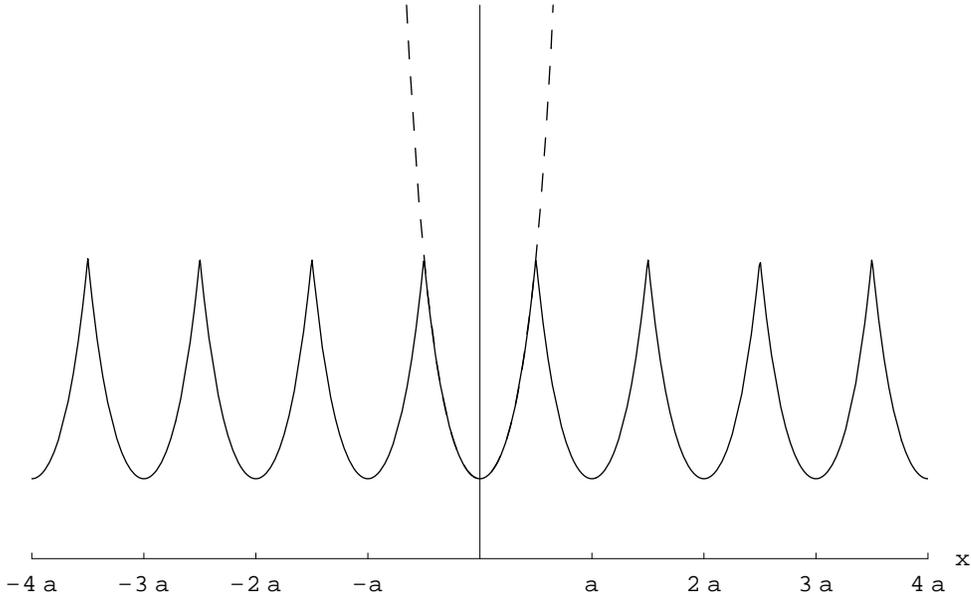}}
\end{picture}
\caption{A graph of $\widetilde G_{array}(X)$,
which has the interpretation of the field produced by an infinite array of $\delta$-function
sources  (`D-branes') located at  $X = a (n +\frac12)$ .
The dashed line represents the analytic continuation to $|X| > \frac{a}{2}$
of the  branch around the origin. }\end{figure}

\smallskip

The advantage of working in
position space is that now we can simply sum over all the
contributions of the array of D-branes explicitly:
\be \label{sumG}
\widetilde G_{array} ( X)
 =
\frac{\pi}{c} \sum_{n=-\infty}^{\infty}
 e^{-c |a(n+\frac{1}{2}) + X|} \; \; .
 \ee
This is a periodic function with period $a$
(see Fig.1),  given in a neighborhood of $X=0$ by
 \be \label{sum}
\widetilde G_{array} ( X) = \frac{\pi \cosh(c X)}{c\sinh(\frac{c
a}{2})}\, , \quad |X| \leq \frac{a}{2} \,. \ee Now we wish to Wick
rotate. Of course, $\widetilde G_{array}(X)$ is {\it not} an
analytic function, precisely because of the $\delta$-function
sources located at $X = (n + \frac12) a$. So we need to specify
what we exactly mean by the analytic continuation $X \to -i X^0$.
A natural   prescription  is to focus on the branch around the
origin to find
\be\label{lj} G_{array} (X^0) =  \frac{\pi \cos(c
X^0)}{c\sinh(\frac{c a}{2})}\, . \ee
Fourier transforming \refb{lj}  back to momentum space we find
 \be
\label{AE} S(E, \dots)= \frac{ \pi}{ 2 c\, \sinh(\frac{c a}{2})}
\,(\delta (E-c)+\delta(E+c)). \ee As anticipated by the heuristic
discussion in the previous subsection, while $\tilde{A}(P, \dots)$
is non-zero for any real  $P$, we find that $S(E, \dots)$ has
support only for those values of $E$ corresponding to
singularities of  $\tilde A(P, \dots)$, namely $E= -i P = \pm c$.
As we shall see in
 the context of string theory this  will have the interpretation
of a change in the dimension of the moduli space.

\subsection{General prescription}

In principle, all amplitudes considered in the paper could be
expanded in terms of the example studied above.  It would be nice
however to have a general formula that gives $S(E,...)$ in terms
of $\widetilde A(P, \dots)$.
To this end
consider, for generic $\widetilde A(P, \dots)$, \be \label{e2}
\widetilde G_{array}(X) = \int_{-\infty}^\infty dP \,e^{i P\, X}
\sum_{n=-\infty}^\infty (-1)^n \, 2 \pi \delta(a P -2 \pi n)
\widetilde{A}(P, \dots) . \ee
In all cases that we study in the present paper $\widetilde{A}(P,
\dots)$ is an analytic function with
 poles or cuts only along the
{\it imaginary} $P$ axis. Moreover $\widetilde{A}(P, \dots)$  goes
to zero for $|P| \to 0$ sufficiently fast  to validate to
following argument.

\smallskip

 With the help of the residues theorem
we can write
\be
\widetilde G_{array}(X) = \frac{1}{2 i }\, \oint_{{\cal C}}dP \, e^{i P\,  X}
\frac{\widetilde{A}(P, \dots)}{\sin(\frac{aP}{2})}\,,
\ee
where the contour ${\cal C}$ is depicted in Fig.2.
 By our analyticity assumptions on $\widetilde{A}(P, \dots)$, the
curve ${\cal C}$ can be deformed to $\tilde{{\cal C}}$ without
crossing any singularities (see Fig.2) and without picking up any
contributions from the two semi-circles at infinity (that are not
shown in the figure). So we conclude, after Fourier transforming
back to momentum space, that
\be\label{disc} S(E) = F (E) \, {\rm Disc}_E[{ \widetilde A(i E) }] \, \ee
where
\be\label{Farray} F(E)  = \frac{1}{2 \sinh \left( \frac{a E}{2}
\right)} \, .  \ee
 Here by ${\rm Disc}_E$ we mean the discontinuity with respect
to $E$, namely \be {\rm Disc}_{E}[f(E)] = \frac{ f(E + i \epsilon)
- f(E - i \epsilon)}{i}  \, ,
\ee
so for example  ${\rm Disc}_E
(1/E) = - 2 \pi \delta(E)$.
Let us check  our master  formula
(\ref{disc}, \ref{Farray}) in the example studied in the previous
subsection. From \refb{i3},  $\widetilde A(i E) = 1/(-E^2 + c^2)$;
applying (\ref{disc}, \ref{Farray}) we immediately reproduce the result
\refb{AE}.

\sectiono{Disk amplitudes }

To gain some insight into the meaning
of the D-brane array at imaginary times \refb{Array},
 we now turn to a detailed analysis of disk  scattering amplitudes of
external closed strings.
 We start by considering the simplest possible case, namely the amplitude
of two closed string tachyons.  We first study this concrete
example using standard methods, and then describe a more abstract
point of view, that can be generalized easily to $m$-point
functions involving arbitrary on-shell closed strings. A  clear
physical interpretation will emerge.

\begin{figure}
\begin{picture}(250,242.5)(0,0)
\vspace{-5mm} \hspace{20mm} \mbox{\epsfxsize=130mm
\epsfbox{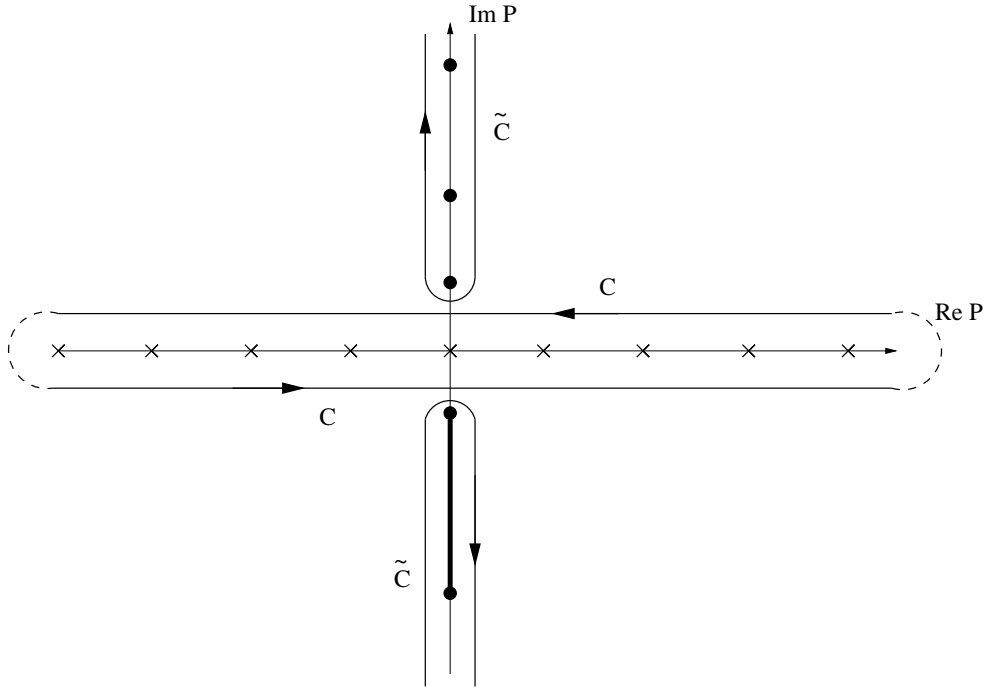}}
\end{picture}
\caption{Integration contours in the complex $P$ plane.
 The zeros of $\sin(a P/2)$ are denoted by the symbols `$x$'
along the real $P$ axis. The black dots represent possible poles
of $\widetilde{A}$ and the thick line represents a possible cut.
We assume that the only singularities of $\widetilde A$ are on the
imaginary $P$ axis.}
\end{figure}

\subsection{Tachyon two-point amplitude}

The simplest non-trivial case that we  consider is the disk
amplitude with the insertion of two closed string tachyons. This
example, which we are going to work out in full detail, contains
already much of the essential physics.

\smallskip

To apply the prescription derived in the previous section, we need
to evaluate the disk two-point function ${\widetilde A}(p_1, p_2)$
for a standard D-brane. We consider a D($p$-1) brane with
Dirichlet boundary conditions for $\widetilde X^M$, $M =0, \cdots
, 25-p$ and Neumann boundary conditions for $\widetilde X^m$,
$m=26-p, \cdots , 25$. We are using a notation that will be
natural for the theory {\it after} the Wick rotation: $\widetilde
X^0$ is a spatial coordinate that will become timelike after Wick
rotation. In order to have a standard D-brane with Neumann
conditions in time, we take one of the $\widetilde X^m$ to be
timelike. So the Wick rotation is actually a double Wick rotation:
It transforms $\widetilde X^0$ from spacelike to timelike and one
of the parallel directions $\widetilde X^m$ from timelike to
spacelike\footnote{The case $p=0$ is special, as we must choose
one of the transverse directions $\widetilde X^M$ to be
timelike.}. In practice, the amplitude $\widetilde A(p_1, p_2)$
will be written in terms of kinematic invariants, so these
distinctions are of no much consequence.

 \smallskip

It is convenient to break up
the momenta into parallel and perpendicular directions,
\be
 p^\mu = (p^M_{\perp},  p^m_{\parallel} ) \, ,
\ee
and define the kinematic invariants
\ben
s & = & p_{1\parallel}^2 = p_{2\parallel}^2  \\
t & = & (p_1 + p_2)^2 = (p_{1\perp} + p_{2 \perp})^2\,.
\nonumber
\een
In our conventions\footnote{Moreover
in writing a string amplitude $\widetilde
A(p_1, \dots p_m)$ , we treat all momenta
as incoming. Finally our convention for the Minkowski metric is
`mostly plus'.}
 $\alpha'=1$, so that the on-shell condition for
closed string tachyons is $p^2 = 4$.

\smallskip

The two-point disk amplitude has one modulus (four real
coordinates minus three conformal Killing vectors).  We work with
a slightly unconventional parametrization of this moduli space
that will be easy to generalize later when we turn to higher-point
functions. Namely we shall fix the positions of the two vertex
operators and integrate over the radius of the disk. We represent
the disk as the complex domain ${\cal H}_{\rho} = \{ z \in{\bf C},
|z| \geq \rho \}$, that is we cut out a hole of radius $\rho$ from
the complex plane. The  measure for the modulus $\rho$ is \be (b_0
+ \bar b_0) \frac{d\rho}{\rho} \,. \ee Using the standard doubling
trick, a closed string vertex operator $V(p , z , \bar{z} )$ is
replaced by the two chiral insertions $V_L(p, z)$ and $V_L(p',
\rho^2/\bar z)$, where $p'= (-p_\perp, p_\parallel  )$. For
closed string tachyons, $V(p, z, \bar z) = c(z) \bar c (\bar z)
\exp(i p X(z, \bar z))$ and $V_L(p, z) = c(z) \exp(i p X_L(z))$.
Fixing the tachyon vertex operators at $z =1$ and $z = \infty$, we
have
 \be {\widetilde A}(p_1, p_2) = \frac{1}{2\pi i}\int_{0}^1
\frac{d\rho}{\rho}\,  \oint_{|z| = \rho }
 dz \;
 \langle \, V_L(p_1; \infty) V_L(p_2; 1) \,
 b(z)  \, V_L(p'_2;
\rho^2) V_L(p'_1; 0) \, \rangle \, ,
 \ee
where the symbol $\langle \, , \rangle$ denotes a CFT correlator
on the plane.
A short calculation gives
\be
 {\widetilde A}(p_1, p_2) = \int_0^1 d\rho \, \rho^{t/2 -3}
(1-\rho^2)^{s-2} \, = \frac{\Gamma({t}/{4} -1) \Gamma(s-1)}{2 \,
\Gamma(t/4 + s -2)} \,.
\ee
This is of course a standard result, with a familiar
interpretation \cite{kt,hk}. The scattering amplitude of a closed
string off a D-brane shows the usual `dual' structure with poles
both in the open and in the closed string channel. In the open
string channel, the poles are located at $s= 1, 0, -1, \dots$ and
arise from expanding around  $\rho = 1$ where the vertex operators
approach the boundary. In the closed string channel, we see poles
at $t = 4, 0, -4, \dots$, arising from expanding around  $\rho =
0$ where the  boundary shrinks to zero size.

\smallskip

We are now in a position to apply the prescription
\refb{disc}\footnote{Notice that ${\widetilde A}(p_1, p_2)$ obeys our
analyticity assumptions, indeed it is an analytic function of the
total momentum $P^0$ with singularities (poles) only for imaginary $P^0$,
and
behaves as $1/|P^0|^{2s}$ for large $|P^0|$.}.
The discontinuity with
respect to $E = E_1 + E_2 = -i (p^0_1 + p^0_2)$
comes from the poles in the variable
$t$, and so we find that
 \be \label{2t} S(p_1, p_2) =
\frac{1}{2 \sinh\left( \frac{a |E|}{2}\right)}\; \sum_{k = 0}^\infty f_k(s)
\, \delta(t/4 -1 +k)\, , \ee where \be \label{Fk} f_k(s) =
\frac{(-1)^k \Gamma(s-1)}{2 \,  k! \Gamma(s -k-1)} = \frac{(2-s)(3-s)
\cdots (1+k -s)}{2\,  k!} \,. \ee

\bigskip

Several remarks are in order. First, all the contributions to $S$
come from the $\rho \to 0$ region of the moduli space. The
sharpest way to see this is to introduce a cut-off $\epsilon \leq
\rho \leq 1$. Then the amplitude $\widetilde A$ becomes analytic
in $t$ since the poles in the closed string channel disappear.
Applying \refb{disc} yields $S=0$ for any $\epsilon $.  This
vindicates the original intuition that the boundary state for the
array of imaginary D-branes should correspond to a `hole of zero
size' (an extra puncture) in the worldsheet. This intuition will
be made very precise below. Second, $S$ has no  poles in $s$, the
open string channel! The external closed strings do not couple to
any on-shell open string degrees of freedom. Since there are no
D-brane sources in real time, this is as expected. We are
describing a {\it purely closed} string background.

\smallskip

More precisely,  \refb{2t} describes a {\it sphere}   amplitude
with  the two tachyons insertions and an additional on-shell
closed string state that involves excitation at all levels of  the
tower of massive modes.
Indeed, the prefactor $f_k(s)$ is a polynomial of
degree $2k$ in $p_{\parallel}$, consistently with the fact that a
closed string mode at level $k$ has up to $2k$ Lorentz indices.

\subsection{Boundary state computation}

In order to secure this result, and to prepare for the
generalization to higher-point amplitudes, we now repeat the
computation in a more abstract language. We still write the
amplitude in the domain ${\cal H}_{\rho}$, but instead of using
the doubling trick, we represent the effect of the boundary at
$|z| =\rho$ by the insertion of a boundary state
$|\widetilde {\cal{B}}^{p-1} \rangle_{|z| = \rho}$. This is the full boundary
state defining the D($p$-1) brane located at $\widetilde X^M=0$.

\smallskip

A boundary state is a ghost number three
 state in the closed string Hilbert space, obeying among other things
the conditions
\be
 \label{QB} (Q_B + \bar
Q_B) |\widetilde {\cal{B}}^{p-1}\rangle = 0 \, , \quad (b_0 - \bar b_0)
|\widetilde {\cal{B}}^{p-1}\rangle = 0 \,.
 \ee
In radial quantization, a state at radius $\rho$ can be obtained
from a state at radius $\rho=1$ by propagation in the closed
string channel,
\be |\widetilde {\cal{B}}^{p-1}\rangle_{|z| = \rho} = \rho^{L_0 + \bar L_0}\,
 |\widetilde {\cal{B}}^{p-1}\rangle_{|z| =1} \,.
\ee
We can then write
\be {\widetilde A}(p_1, p_2) =
 \int_{0}^1 \frac{d\rho}{\rho} \;
\langle | \, V(p_1; \infty, \infty) V(p_2; 1,1) \, (b_0 + \bar b_0)
\; \rho^{L_0 + \bar L_0} \; | \widetilde {\cal{B}}^{p-1} \rangle_{|z| =1} \, \,.
\ee
To proceed, we
insert a complete set of states $\{  |k, i \rangle \}$,
\be
{\widetilde A}(p_1, p_2) = \int dk \, \sum_i \, \int_{0}^1
\frac{d\rho}{\rho} \, \langle | \,V(p_1; \infty, \infty) V(p_2; 1,1)
\, |k, i \rangle \, \rho^{(\frac{k^2}{2} + 2 l_i)}\, \langle k, i |(b_0 + \bar
b_0) |\widetilde {\cal{B}}^{p-1} \rangle_{|z| =1} \,, \ee
where $l_i$ is the level of the state $|k, i \rangle$, and we
integrate over the modulus $\rho$ to get
\be \label{intt} {\widetilde A}(p_1, p_2) = \int dk \, \sum_i \langle |\,
V(p_1; \infty, \infty) V(p_2; 1,1) \, |k, i \rangle \;\frac{1}{
\frac{k^2}{2} + 2 l_i} \;
\langle k, i |(b_0 + \bar b_0) | \widetilde {\cal{B}}^{p-1}
\rangle_{|z| =1} \, . \ee
This expression exhibits the decomposition of the amplitude into a
source term from the boundary state, a closed string propagator,
and  the 3-point interaction vertex. If we now apply
the master formula \refb{disc} we find
 \be \label{correlator} S (p_1, p_2)
= \langle |\, V(p_1; \infty, \infty) \,V(p_2; 1,1)\, |W \rangle\, , \ee
where\footnote{Notice that in
the formula below
we drop the `tilde' on the symbol for the boundary state:
$ |{\cal{B}}^{p-1} \rangle$ denotes the boundary
state {\it after} double Wick rotation.}
\ben\label{67}
 |W \rangle & \equiv &
\int dk \sum_i |k, i \rangle \; \frac{\delta(k^2/2+ 2 l(i))}{2
\sinh\left(\frac{a| E|}{2}\right)} \;
\langle k, i |\, (b_0 + \bar b_0) | {\cal{B}}^{p-1} \rangle_{|z| =1} \\
& = &
\frac{\delta(L_0 + \bar L_0)}{2 \sinh\left(\frac{a| E|}{2}\right)}\,
(b_0 + \bar b_0) |{\cal{B}}^{p-1} \rangle_{|z| =1} \, \nonumber .
\een
We see that the interaction of the two tachyons with the imaginary
array is captured by a ghost number two closed string state $|W
\rangle$. Using \refb{QB} and
$\{ b_0, Q_B \} = L_0$, we have
\be
 (Q_B + \bar Q_B)  |W \rangle = 0 \,, \quad (b_0 - \bar b_0) |W
\rangle \,.
\ee
These are
precisely the physical-state conditions for closed strings.
Moreover, we have the freedom to add to
 $|W \rangle$  BRST trivial states, that would decouple
in the computation of the correlator \refb{correlator}. We
recognize $|W \rangle$ as an element of the closed string
cohomology. A more detailed discussion of this state and of its
space-time interpretation will be carried out in  section 4.

\smallskip

Finally we can  trade the state $|W \rangle$  with a vertex operator
insertion at the origin, and re-write the amplitude \refb{correlator}
as a CFT correlator in the plane,
\be
 S(p_1, p_2) = \langle \, V(p_1; \infty, \infty)\, V(p_2; 1,1)\, {\cal
W} (0,0) \, \rangle\, .
\ee
This is manifestly the scattering
amplitude for three closed strings on the {\it sphere}! Having
explicitly performed the integral over $\rho$
there are no remaining moduli, as it should  be for a {\it sphere}
three-point function.

\smallskip

Clearly this conclusion does not depend on the vertex operators
$V(p_i, z, \bar z)$ being closed string tachyons. The analysis
immediately generalizes to arbitrary physical closed string vertex
operators: in the background of the imaginary array, a two-point
function on the disk is {\it equal} to a three-point function on
the sphere, with the on-shell  vertex operator ${\cal W}$ as the
extra insertion.

\subsection{Higher-point disk amplitudes}

The computation of higher-point amplitudes is now a straightforward
generalization. We still represent the disk as the complex domain
${\cal H}_{\rho}$ and describe the moduli space of $m$ closed
strings on the disk by fixing the positions of two vertex
operators, say $V_1$ at $z=\infty$ and $V_2$ at $z=1$, and varying
the positions of the other $m-2$ insertions and the radius $\rho$
of the hole. More precisely, we vary the $m-2$ coordinates over
the full complex plane, $\{ z_i \in {\bf C}$, $i=3, \dots m \}$, and for
a given choice of the $\{ z_i \}$ we vary the radius $\rho$
between 0 and the distance of the closest insertion, $0 \leq \rho
\leq \rho_0 ={\rm min}[|z_i|, i=1, \dots m]$ (see Fig.3).
This way we cover moduli space exactly once,
as can be easily checked for example
by mapping the above configuration to the the interior of the unit disk.
 We then have
\ben
&& {\widetilde A} (p_1,
\dots, p_n) = \int d^2z_3  \dots d^2z_m\,
\int_{0}^{\rho_0} \frac{d\rho}{\rho} \, \\
&& \langle |\, {\cal R}\{ \,V_1(p_1; \infty, \infty) V_2(p_2; 1,1)
\dots V_m(p_m; z_m, \bar z_m) \}\, (b_0 + \bar b_0) \, \rho^{L_0 +
\bar L_0} \, | \widetilde {\cal{B}}^{p-1} \rangle_{|z| =1} \, \,, \nonumber
\een where  ${\cal R}\{\, . \}$ denotes radial ordering. Inserting
as before an intermediate complete set of states, and performing
the $\rho$ integral, we find \ben && {\widetilde A} (p_1, \dots, p_n)  =\int
dk  \sum_i \int d^2z_3
\dots d^2z_m \\
&& \langle |\, {\cal R}\{ V_1(p_1; \infty, \infty)
V_2(p_2; 1,1) \dots
V_m(p_m; z_m, \bar z_m) \}  | k, i \rangle \nonumber
\frac{\rho_0^{\frac{k^2}{2} + 2 l_i }}{\frac{ k^2}{2} + 2 l_i}  \langle k, i| \,
(b_0 + \bar b_0) \,
 | \widetilde {\cal{B}}^{p-1} \rangle_{|z| =1} \, \,.
\een
Now we extract the discontinuity with respect to the energy $E$.
Clearly we get contributions from the poles in the propagators
$\sim \frac{1}{k^2/2 + 2 l_i}$. A priori, the integrals over the
coordinates $z_i$ may generate additional singularities. However
it is not difficult to show that extra singularities can only
arise when a (proper) subset of the vertex operators have an
on-shell total momentum. We can define the amplitude by analytic
continuation away from these singular points, and then disregard
these singularities. This  is reminiscent of the celebrated
canceled propagator argument \cite{polchinski}. We can then
write\footnote{ Notice that the dependence on $\rho_0$ drops
because of the delta function $\delta (L_0 + \bar{L}_0)$. In other
terms, the $\rho$ integral localizes to $ \rho \to 0$ and the
upper limit of integration $\rho_0$ is immaterial.}(see Fig.3)
\be
 S(p_1, \cdots, p_m) = \int d^2z_3  \dots d^2z_m \, \langle
\, V_1(p_1; \infty, \infty) V_2(p_2; 1,1) \dots V_m(p_m; z_m, \bar
z_m) \} {\cal W}(0,0)\, \rangle\, \,. \ee The coordinates $z_i$,
$i=3, \dots m$, are integrated over the full complex plane, so
this is the string theory amplitude for $m+1$ insertions on the
sphere. We can check that the counting of moduli is consistent
with this result. We started with $2m - 3$ moduli for a disk with
$m$ closed punctures and performed an explicit integration over
$\rho$. This gives $2m -4 =2(m+1) -6$, which is the number of
moduli for a sphere with $m+1$ punctures.

\begin{figure}
\begin{picture}(150,220)(0,0)
 \vspace{-55mm} \hspace{-3mm} \mbox{\epsfxsize=80mm \epsfbox{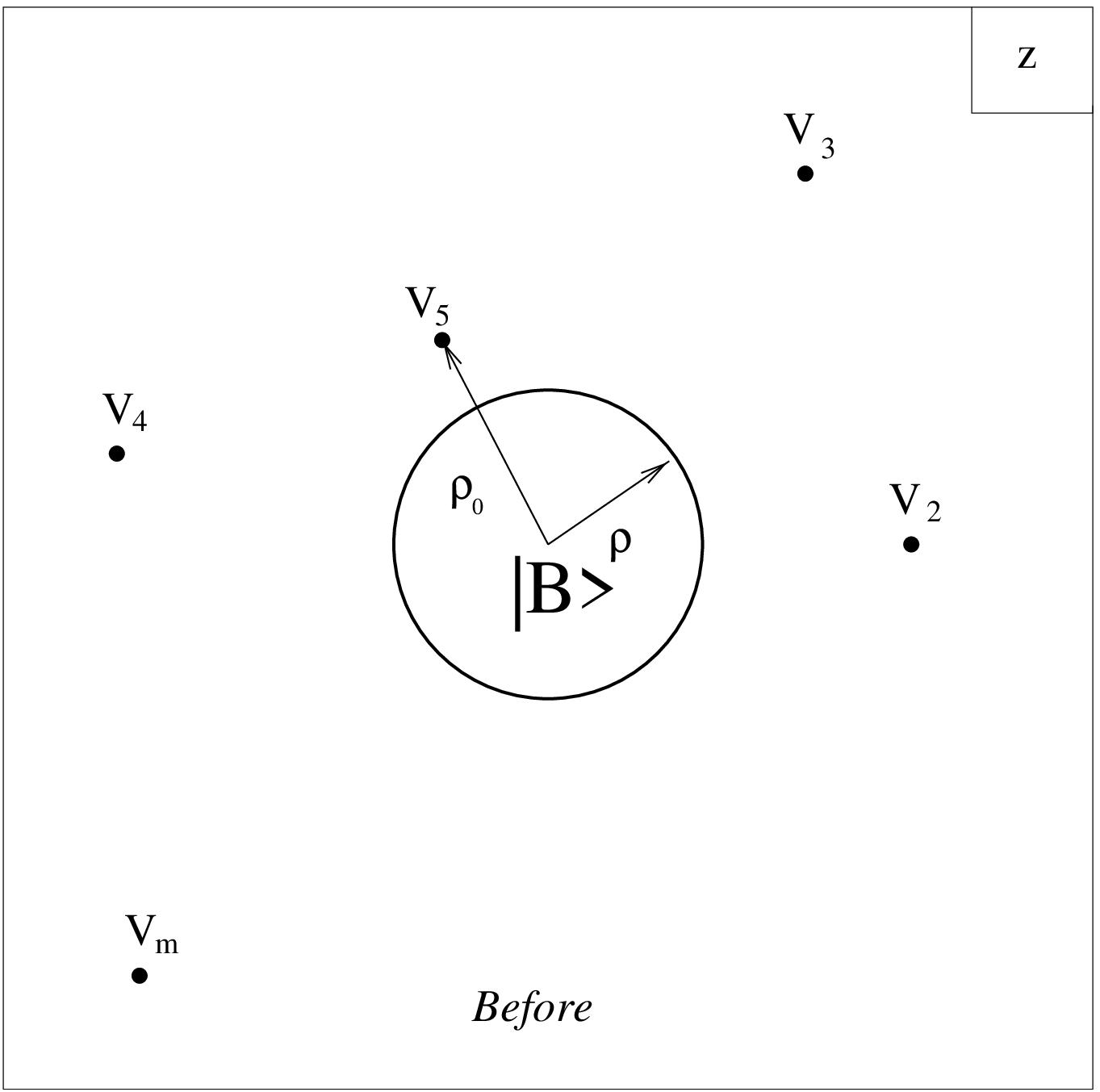}}\mbox{\epsfxsize=80mm
\epsfbox{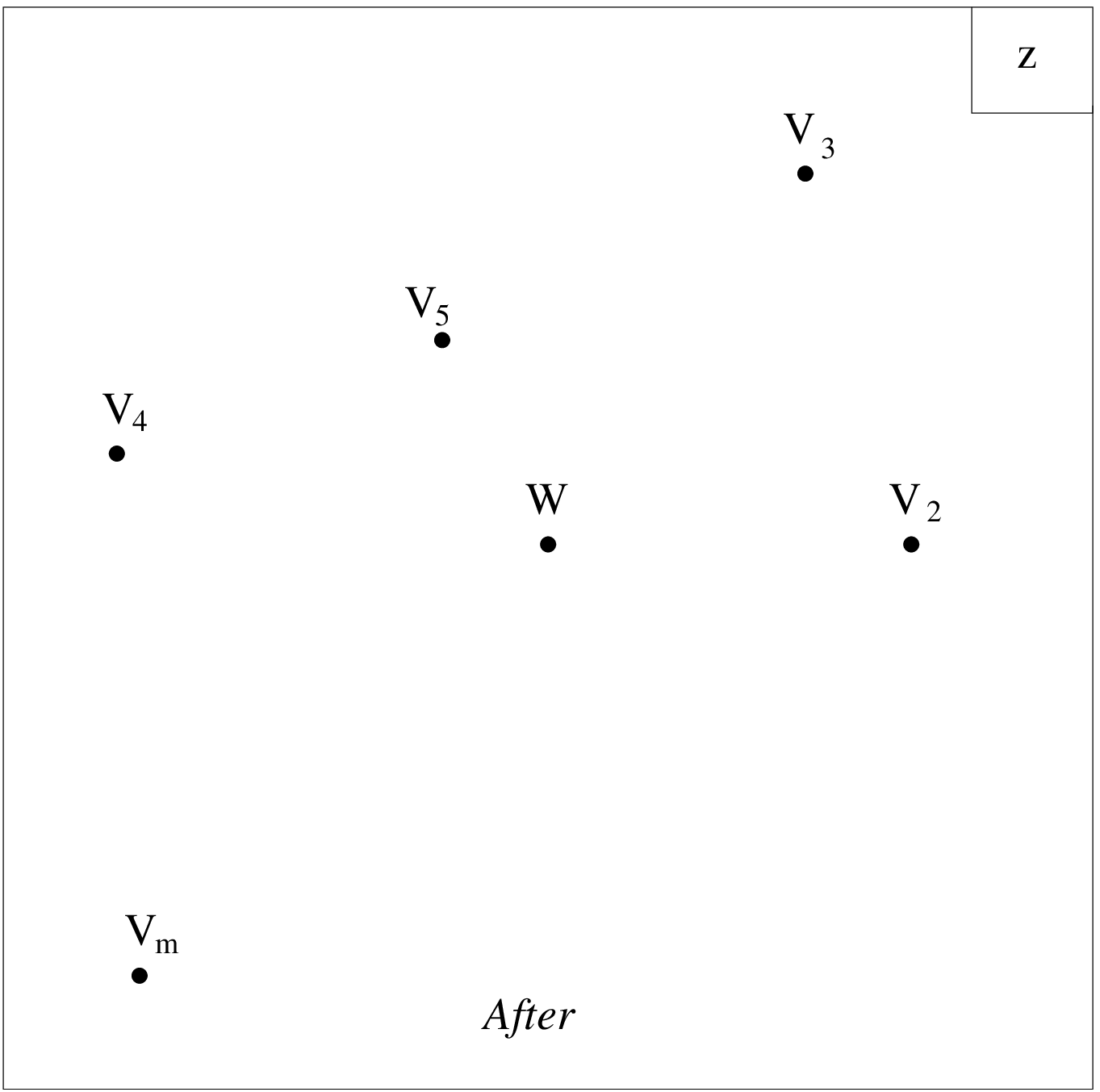}}
\end{picture}
\caption{ Before the double Wick rotation
 we have a standard disk amplitude. The disk can be viewed
as the region $\HH_\rho$, which is the complex plane with
 a hole of radius $\rho$.  There are contributions to the scattering amplitude from all values of
$\rho \leq \rho_0$, where $\rho_0$ is the distance of the closest
puncture. After the double Wick rotation
 the only contribution is coming
from $\rho=0$. The hole shrinks  to a point leaving behind an
extra puncture ${\cal W}$ inserted at the origin.}
\end{figure}

\subsection{Interpretation}

Let us summarize what we have learned. {\it  Disk amplitudes for
{\rm m} external closed strings off the D-brane array at the
imaginary times $X^0 = i (n +1/2)a$ are completely equivalent to
{\rm m+1}-point amplitudes on the {\rm sphere}, with the extra
closed string insertion} $$g_s \int \, d^2z \,  {\cal W}(z, \bar
z) \, ,$$ {\it where} ${\cal W}$ {\it is given by} \refb{67}. The
explicit factor of $g_s$, which was omitted in the previous
formulae, comes from the relative normalization of disk amplitudes
to standard sphere amplitudes.  Unlike the case of an ordinary
brane, we are finding that for the imaginary array  disk
amplitudes give a {\it purely closed} correction (of order $g_s$)
to the background. There are no D-brane sources in real time, only
closed strings satisfying the {\it homogeneous} wave
equation.


The absence of sources in real time can also be deduced from the
fact that all {\it one}-point functions on the disk are trivially
zero in our prescription.  This is consistent with the
computations in \cite{s}, where the stress tensor (related to the
graviton one-point function) was found to vanish in the BCFT
\refb{iu} for $\lambda =\frac12$.
There are two equivalent ways to see this in our language. The
one-point function on the disk is a smooth function of the total
energy $E$, so there is no discontinuity in $E$. Alternatively,
the discussion above implies that a disk one-point function is
equal to a sphere {\it two}-point function with the extra
insertion of ${\cal W}$; but sphere two-point functions are zero
because of the infinite volume of the unfixed moduli.

\smallskip

Since disk amplitudes provide the first order correction to the background,
it is very natural to expect that amplitudes with multiple
boundaries give the necessary higher-order corrections.
A genus zero amplitude with $m$ external closed
strings and $b$ boundaries,  is expected to become
after Wick rotation
a sphere amplitude with $m +b$ punctures, with $b$
 insertions of $g_s \int d^2 z {\cal W}$. Notice that  since
the boundaries are indistinguishable, the sum over boundaries {\it
exponentiates} to the insertion of $  \exp ( \int \,d^2 z \, g_s
{\cal W})$, which has the interpretation of a coherent state of
closed strings.
 One may investigate this issue rigorously by representing the effect of the boundaries
using boundary states,  and  decomposing the moduli space of a
surface with $b$ boundaries and $m$ closed string punctures in
terms of closed string vertices and propagators \cite{zwi2}. The
moduli space integration should localize to the region where each
boundary shrinks to an extra puncture.
 A priori, one may also expect that
regions of moduli space where zero-size boundaries  collide could
provide additional contributions. For example one may expect that
an annulus amplitude would reduce to a sphere amplitude with two
insertions of $g_s {\cal W }$, {\it plus} a sphere amplitude with
a single insertion of a new operator $g_s^2 {\cal W}_1$ coming
from the shrinking of both boundaries to the same point. This
issue should be investigated further.

\sectiono{The closed string state}\label{aa}

In this section we  study the physical properties of $| { W} \rangle$
in more detail and make contact with \cite{strom} and
\cite{mald}.
Our starting point is the boundary state $|{\cal B} ^{p-1}\rangle$,
which is obtained from double Wick rotation of $|\widetilde
{\cal B} ^{p-1}\rangle$. So  $|{\cal B} ^{p-1}\rangle$ is
associated with a D($p$-1) brane
located at  $X^M =0$, in a space with metric $(\eta_{MN}, \delta_{mn})$.
One has
\be
|{\cal B} ^{p-1}\rangle = {\cal N} \delta(X^M)\exp
\left(
-\sum_{n=1}^{\infty} a_n^{\mu\dag}{S}_{\mu\nu}
\tilde{a}_n^{\nu\dag}\right) |0;k=0\rangle \, . \ee
${\cal N}$ is a normalization constant that can be found for
example in \cite{068}, and
\be
\label{456}
S_{MN}=-\eta_{MN},\;\;S_{mn}=\delta_{mn}.
\ee
With the help of (\ref{67}) we can expand  $| { W} \rangle$
in terms of  closed string physical states,
\[ | { W} \rangle ={{\cal N}} \, c_1 \bar c_1 | 0 \rangle_{gh}
\otimes \int \, \frac{dk^{25-p}_{\perp} }{(2 \pi)^{25-p}}
\frac{1}{{2 |E|}} \frac{1}{2 \sinh\left(\frac{a |E|}{2}\right)} \,
\Big( | k^2 = 4 \rangle -   S_{\mu \nu} \, \partial X^\mu \bar
\partial X^\nu | k^2 = 0 \rangle + \cdots \Big) \, ,\] where the
dots indicate higher massive modes. The first term is the closed
string tachyon, which is an artifact of the bosonic string and is
standard practice to ignore. Below we first discuss the massless
modes and then consider the massive modes.

\subsection{Massless sector}

In order to read off the dilaton and gravity wave profile from the
second term we have to undo their mixing. The unmixed dilaton and
graviton (in the Einstein frame) take the form (see {\it e.g.}
\cite{068}) \be \label{yy} h_{\mu\nu}= S_{\mu\nu}-\frac{S\cdot
\epsilon^{(\phi)}}{\eta \cdot \epsilon ^{(\phi)}}\;
\eta_{\mu\nu},\;\;\;\;\phi=S\cdot \epsilon^{(\phi)}, \ee where \be
\epsilon^{(\phi)}=\frac{1}{2}
(\eta_{\mu\nu}-k_{\mu}l_{\nu}-k_{\nu}l_{\mu})\, ,\;\;\;\;k\cdot
l=1, \;\;\;l^2=0. \ee Let us first look at the case $p=0$ (an
array of imaginary D(-1) branes). One finds
 that the Einstein metric is completely flat as all the
expectation values of $h_{\mu\nu}$ vanish. This means that  the
part of the leading term in the ADM mass,  which scales like
$1/g_s$, vanishes  (see also \cite{mald})\footnote{This is
consistent with Sen's observation \cite{s} that the stress tensor
vanishes for the $\lambda = \frac12$ BCFT \refb{iu}.}. The
dilaton, on the other hand, does not vanish. So for $p =0$, the
massless fields in $|{ W} \rangle$ consist of a spherically
symmetric dilaton wave $\phi(r, X^0)$ in 25+1 dimensions, whose
energy is of order $O(g_s^0)=O(1)$.

\smallskip

For $p>0$ the 26-dimensional metric in the Einstein frame is
non-trivial. However the fields profiles are translationally
invariant in the $p$ longitudinal directions $X^m$, and to read
off the ADM mass we are instructed to  dimensionally reduce  to
the $26-p$ transverse dimensions. One finds  that in the
$26-p$-dimensional Einstein frame the metric {\it is} zero, and we
have again only a spherically-symmetric dilaton wave. We  conclude
that  to order $1/g_s$ the mass vanishes for general $p$.

\smallskip

The space-time profile of this dilaton wave is quite interesting
 \cite{strom}.\footnote{From the discussion in section 6 of
\cite{strom} it is not obvious that eq.(6.27) in that section
describes just a dilaton wave. However, this can be verified with
the help of  eq. (\ref{yy}) above.} For fixed radius $r = \sqrt{
X^M X^M}$,  the field decays exponentially fast as $X^0 \to \pm
\infty$. For fixed $X^0$, the field decays as $1/r^{23-p}$ as $r
\to \infty$, just like the fields produced by an ordinary
D$p$-brane, but with a different numerical coefficient. (These two
asymptotic behaviors match in a region of thickness of the order
of  $a$ around the light-cone $X^0 = \pm r$). To be precise, up to
an overall numerical constant that will not  be relevant for the
discussion below, the leading asymptotic behavior as $r \to
\infty$ of the various fields is, in the 26-dimensional Einstein
frame, \ben \label{s1} && h_{00} \to \frac{(1+K)(d-3)-2p}{d-2} \,
\frac{1}{r^{23-p}} ,\nonumber
\\
&& h_{mn} \to \frac{\delta_{mn}}{2}\left(\frac{2d-2p-5-K }{d-2}\right) \, \frac{1}{r^{23-p}},
\\
&& h_{MN} \to \frac{\delta_{MN}}{2}\left(\frac{2p+1+K}{d-2}\right) \, \frac{1}{r^{23-p}}, \nonumber
\\
&& \phi \to \frac{d-2p-3-K}{4} \, \frac{1}{r^{23-p}}\, .\nonumber
\een Here $d = 26$ and the parameter $K$ is set to $1$ for
the standard D$p$-brane and
is set to $-1$ for the  background $|W \rangle$.

\smallskip

An  exercise  one can now do is to compute the force acting on a
probe D0-brane in this dilaton wave background \cite{strom}. One
has to be a bit careful here with the exact meaning of the `force'
between the brane and the wave since this is not a static set-up.
The `force' that can be computed using \refb{s1} for $K=-1$ is
acting during a finite time interval $ -T \leq X^0\leq T
$ at a distance $r$ in the limit $\frac{T}{r} \rightarrow
0$. In the Einstein frame in 26 dimensions the DBI action for the
D0 brane takes the form \be \label{9w1} \int dX^0 \,
e^{-\frac{11}{12}\phi} \sqrt{g_{00}}\, . \ee
From this and \refb{s1} with $p=0$
we can
deduce that the ratio between the force acted upon the D0 by a
standard brane and the `force' acted on it by the dilatonic wave
in $|W \rangle$ is \be
\frac{F_{standard}}{F_{W}}=\frac{\frac{11}{12} \phi(K=1) +\frac12
h_{00}(K=1)}{\frac{11}{12} \phi(K=-1 ) +\frac12 h_{00}(K=-1)}
=\frac{12}{11}. \ee This is in agreement with \cite{strom} where
this ratio was calculated in a different way.

\subsection{Massive modes}\label{energy}

The massive closed string states in $| W \rangle$ obey the
dispersion relation $E^2 = k_\perp^2 + m^2$, with $m^2 = 4 (n-1)$.
For $n  > 1$, their field profile (proportional to $1/\sinh(a
|E|/2$))  is strongly peaked at energies $|E| - m \leq 1/a$. This
means that for any  $a \geq 2 \pi$, the states are with  good
approximation {\it non-relativistic} already at level $n =2$, with
the approximation improving at higher levels.
Their fields at $X^0 = 0$ are Gaussians of the form \be
 \delta(k_\parallel) \,  \exp(- \frac{ m a}{2} )  \exp(-\frac{a k_\perp^2}{4 m}) \,  ,
\ee
that is, the closed string modes occupy the directions
$X^m$ and are localized
 at $X^M = 0$ ($M \neq 0$)
with a width of order
$\sqrt{ a/m}$ in the transverse directions.
 The time evolution of these field profiles follows
non-relativistic Schr\"odinger equation,
so their width scales as $|X^0|/ \sqrt{a m}$ for
large times.  Interestingly, the massive modes behave
as non-relativistic matter located at the would-be position
of the brane.  This conclusion was also reached in \cite{mald}.


Let us now compute the normalization
of $|W \rangle$ and its the space-time energy,
following \cite{mald}\footnote{Equs. \refb{barE} and
\refb{Ep} below were {\it not} obtained independently of \cite{mald}.}.
The normalization has the interpretation of
(the expectation value of) the total number of particles  $\bar n$
in the background.
One has
\ben \label{barE}
\bar n =  \langle {\cal W} | c_0 \bar{c}_0 |{\cal W }\rangle
& =&
{\cal N}^2\,
\sum_{n=0}^{\infty}
d_n\, \int\frac{dk_{\perp}^{25-p}}{(2 \pi)^{25-p}}\frac{1}{2|E|}\, \frac{1}{4\sinh^2 \left(\frac{aE}{2}\right)}\, ,
\\
\bar E & =&{\cal N}^2\,
\sum_{n=0}^{\infty}
d_n\, \int\frac{dk_{\perp}^{25-p}}{(2 \pi)^{25-p}}\frac{1}{2|E|}\, \frac{|E|}{4\sinh^2 \left(\frac{aE}{2}\right)}\, ,
\nonumber \een
where $d_n$  can be computed from the generating function
\be \sum_{n=0}^{\infty} d_n
w^n=f(w)^{-24},\;\;\;\;f(w)=\prod_{m=1}^{\infty} (1-w^n).
\ee
The asymptotic behavior of $d_n$ for large $n$ is
\cite{gsw} $ d_n\sim n^{-27/4}e^{4\pi \sqrt{n}} \, .
$
It is easy to see that for $a  > 2\pi$ the exponential suppression from
$1/\sinh^2(\frac{a|E|}{2})$ wins over the exponential growth of states,
and both $\bar n$ and $\bar E$
are perfectly {\it finite}.
On the other hand,  in the limit $ a \to 2 \pi$ we get
 \be
\label{Ep} \bar E \sim \sum_{n=0}^\infty \frac{ e^{- 2 (a - 2 \pi)
\sqrt{n}}
 }{ n^{p/4 }}
\sim  (a - 2 \pi)^{p/2-1} \, .
\ee
For $ a = 2 \pi$ the expectation value of the energy diverges for
$p=0,1,2$.
 Naively, for $p>2$ the energy is finite. However,   for any $p$  the expectation values of powers $\langle E^k
\rangle$ will eventually diverge for sufficiently high $k$
\cite{mald}, and hence  for $a=2\pi$ the uncertainty in the
energy\footnote{A similar behavior is found for $\bar n$, which is
logarithmically divergent for $p=0$ and finite for $p > 0$, but
with higher moments diverging for any $p$.} is infinite for any
$p$.

\smallskip

This divergence has a very natural physical interpretation. As
reviewed in the introduction and further discussed in section 8
below, for $a = 2 \pi$ the background of imaginary branes admits
an {\it infinitesimal} deformation that introduces a D-brane
source in real time, and we should then expect that this
background also has an energy of order $1/g_s$. Since we are
computing this energy in the limit $g_s \to 0$ in a perturbative
expansion $\bar E = \sum_{n=0}\, (g_s)^n\, E^{(n)}$, it is natural
to find that leading term $E^{(0)} = \infty$. At finite $g_s$ this
divergence should be regulated in such a way that
 $\bar E \sim 1/g_s$.

\smallskip
Before discussing a heuristic mechanism that supports this
expectation, we would like to point out that for $a = 2 \pi$ the
state $| W \rangle$ has all the features to be {\it identified}
with `tachyon matter' \cite{s, 4}. Indeed for $a = 2\pi$ the energy is stored
in very massive closed strings modes that behave like
non-relativistic matter strongly localized along the directions
$X^m$.
This is  the closed string dual of Sen's discussion on `tachyon
matter' in the context of {\it open} string field theory; for
large times $|X^0| \to \infty$ the classical open string solution
corresponding to the BCFT \refb{iu} approaches, for all $\lambda$,
an open string configuration
 with zero pressure and with all the energy localized along the $X^m$
directions\footnote{
More precisely, as shown in \cite{mald}
and further elaborated in section 8 of this paper,
the limit $X^0 \to \infty$ of \refb{iu} corresponds for all $\lambda$
(up to a trivial time translation) to the {\it outgoing} ($X^0 > 0$)
part of our state $| W \rangle$, and symmetrically
$X^0 \to -\infty$ of \refb{iu} corresponds to the {\it incoming}
$(X^0 < 0)$ part of  $|W \rangle$.}.

\smallskip

It is very tempting to suspect that at finite
$g_s$ the distance $a$  is renormalized to an effective value
\be\label{45} a_{eff}=2\pi+\varepsilon
\;\;\;\mbox{where}\;\;\;\varepsilon= \gamma \, g_s^b \, ,\ee where
both $b$ and $\gamma$ are positive numbers of order one. This
would make the energy finite. Restricting in the following to $p =
0$\footnote{ For  $p>0$ \refb{Ep} cannot be trusted as the higher
moments of the energy $ ( \langle E^k \rangle - \langle E
\rangle^k)^{1/k}$
 have a worse degree of divergence than the mean value $\bar E = \langle E \rangle$.},
 we find from \refb{Ep}
\be\label{3t2} \bar  E\sim \frac{1}{\varepsilon}.
\ee
So to obtain the expected scaling $\sim 1/g_s$ the parameter
 $b$ has  to be equal to one. An argument why this is plausible
was given in \cite{mald} from the point of view of {\it open} string
field theory. Here  we provide an alternative heuristic
argument  from the {\it closed} string channel,
that justifies why $b=1$ and also why $\gamma > 0$.

\smallskip

Let us think about this issue from the point of view
of the array before Wick rotation. When $g_s=0$ the distance between the
branes is $2\pi$. When $g_s$ is turned on the branes will slightly
curve space-time due to their mass so that the distance between
them is no longer $2\pi$. Actually since their mass is $\sim 1/g_s$
and the Newton constant scales like $g_s^2 $ the back-reaction is of
order $g_s$ and so the proper distance between the branes is indeed
as in (\ref{45}). $\gamma$ is positive simply because the
metric components in the transverse direction to the brane scale
like
\be
g_{\perp} \sim 1+g_s \, \frac{\tilde{\gamma }}{r^{23-p}},
\ee
 with $\tilde{\gamma}> 0$.

\smallskip

It is interesting to check if this argument
 generalizes when we put $N$ rather than just one brane at each site.
 First, the back-reaction of the metric  scales like $g_s N$ which means
 that now $\varepsilon \sim g_s N$.
Combining this with (\ref{3t2}) and with the fact that $| W \rangle$
gets multiplied by a factor of $N$ we
find for the total energy
\be \bar E_N \sim \frac{N^2}{\varepsilon} \sim \frac{N}{g_s},
\ee
which is the correct scaling for the tension of $N$ D-branes.

\smallskip

Clearly this heuristic reasoning is not powerful enough
 to fix $\gamma$ in (\ref{45}). One obvious reason is that the
 distance between the branes is of the order of the string scale
 and hence the gravity approximation should not be trusted.
 A closely related point is the following. Suppose that we could somehow  fix
 $\gamma$ and that we found  the correct D-brane mass.
 If $\gamma$ was  just a fixed number then it is easy to see that the {\it uncertainty} in the D-brane
 mass would be of the same order as the mass itself, which is of course
 not the case for D-branes at weak coupling.
This seems to
 suggest  that $\gamma$ should not  be viewed as a constant but rather as
  a fluctuating field.

\sectiono{On open and closed string moduli}

\label{open}

In section 3 we showed that  scattering amplitudes
off the imaginary array do not have
any open string poles. This means that there are no
propagating open strings degrees of freedom,
consistently with the fact that there are no branes
in real time.  However there
still is a discrete set of on-shell open string
vertex operators, which demand an interpretation.
 In this section we argue that they
are dual to deformations of the
closed string state $| W \rangle$.

\smallskip

For simplicity let us start by considering the case $p=0$, the
array of D(-1) branes at imaginary times $X^0 =i  (n+ \frac 12)
a$. The open string spectrum should be read off from the theory
before double Wick rotation, where we have an array of D(-1)
branes at the spatial locations $\widetilde X^0 = (n + \frac 12)
a$. For generic\footnote{The case $a = 2 \pi$ is of course special
and will be discussed in section 8.} distance $a$, the only matter
primaries of dimension one are $ \widetilde V^{(n)}_{ \mu}=\partial
\widetilde X^{(n)}_{\mu}$, where $\mu$ are space-time indices and
$n \in {\bf Z}$ labels the position of the D(-1) brane. In the
double-Wick rotated theory the physical open string states are
just the same, up to trivial relabeling. So the most general state
in the open string cohomology of the theory of D(-1) branes at
imaginary times can be written as \be \label{oc} | f \rangle =
\sum_{\mu=0}^{25} \sum_{n = -\infty}^\infty f_{(n)}^\mu \,
\partial X^{(n)}_\mu  (0) \, c_1 | 0 \rangle \,.
\ee
Clearly these are the exactly marginal open string deformations
that correspond to moving the positions of the D(-1) branes.

\smallskip
Next we can consider disk amplitudes $S(p_1 \dots p_m; f )$ for
$m$ closed strings scattering off the imaginary array, with one
additional insertion of \refb{oc} on the boundary of the disk.
Without any open strings puncture, $S(p_1, \dots p_m)$ was shown
in section 3 to be a {\it sphere} amplitude with an extra
insertion of the closed string state $| W \rangle$. How does the
addition of $|f \rangle$
 change this conclusion?

\smallskip

To get some insight, consider first the case
$f^\mu_{(n)} = f^\mu$ for all $n$. This deformation
is simply a Goldstone mode associated with a rigid
translation $ X_\mu \to X_\mu + f_\mu$ of the whole
array.  It is clear that in this simple case
$
S(p_1 \dots p_m; f) $   is obtained from  $S(p_1 \dots p_m; f )$ by replacing $| W \rangle$
with its infinitesimal translation, $|W \rangle \to
 f^\mu  \partial_\mu | W \rangle$. In this special example
the open string insertion $|f \rangle$ corresponds
to a symmetry of the vacuum broken by the
closed string background $| W \rangle$.

\smallskip

In the more general case \refb{oc}, it is not difficult to
show\footnote{ We cannot directly apply the prescription of
section 2. However, the relevant disk amplitude is simple enough
that can be  evaluated directly in the theory of the spatial array
$\widetilde X^0 = a(n + \frac 12)$, and then Wick rotated.} that
$S(p_1 \dots p_m; f) $ will still be a sphere amplitude with $m+1$
punctures, where the extra closed puncture is now  $\delta_f {\cal
W}$, the infinitesimal deformation of ${\cal W}$ obtained by
displacing the branes according to $| f \rangle$. Generically this
deformation will change the total energy of the closed string
state. The intuitive picture is that the branes act as sources for
the closed string fields. Since these sources are not in real
time, the resulting closed string fields are {\it homogeneous}
solutions of the wave equation, and we have a purely closed
background. Changing the positions of the branes changes the
details of the closed string field profiles.
 Thus the open string moduli \refb{oc} are re-interpreted as deformations of the closed string
background.   In  the next section  this picture is
made precise.

\smallskip


There is one important restriction on the allowed motions of the
branes:  The resulting closed string fields must be  {\it real}.
This imposes certain constraints on the moduli $f^\mu_{(n)}$.
Starting from the array at $X^0 = i n (a + \frac 12)$, reality of
the closed string fields demands that the D(-1) branes be moved in
{\it pairs}, that is, the $k$-th D(-1) brane at $X^0= i a (k  +
\frac 12) $, $k > 0$,
 together with its mirror partner at $X^0 = i a (-k + \frac 12)$. The reality condition is
\be \label{reality} f_{(k)}^\mu = ( f_{(-k)}^\mu)^*    \, \quad
\forall k \in {\bf N} \, , \ee where $*$ denotes complex
conjugation. As long as we refrain from considering branes with
complex {\it spatial} coordinates $X^i$, $i =1, \dots 25$, we
should keep the spatial moduli $f^i_{(k)}$ real, and then
\refb{reality} implies $f^i_{(k)} = f^i_{(-k)}$. For the time
coordinate $X^0$ on the other hand, there are two possible motions
for a given pair of branes, as  illustrated in Fig.4\footnote{
Interestingly, for a pair of branes before the double Wick
rotation there are also {\it two} allowed exactly marginal motions
along $X = - i X^0$, but of course they  are the independent
translations of each brane along the real $X$ axis.}.

\begin{figure}
\begin{picture}(150,180)(0,0)
\vspace{0mm} \hspace{35mm}\mbox{\epsfxsize=80mm \epsfbox{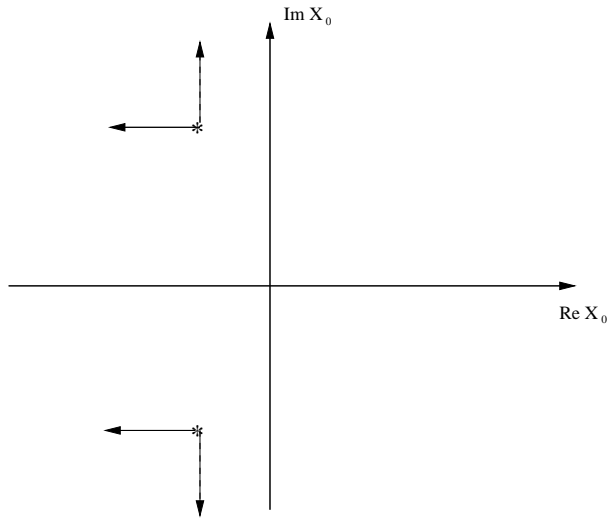}}
\end{picture}
\caption{The two possible motion modes of a pair of D-branes in
the complex $X^0$ plane.}
\end{figure}

\smallskip

The reality condition can be simply rephrased by saying that the
D-brane configuration must be symmetric under reflection with
respect to real time axis.  A simple way to see this constraint is
to focus for example on the massless closed string fields.  The
field produced by some $\delta$-function sources located at the
origin in space and at times $X^0 = Y_k^0$ is schematically
\cite{strom}
\be
\phi (X^\mu)  \sim \sum_k  \frac{1}{[- (X^0 - Y_k^0)^2 +
(X^i)^2]^\sigma} \ee
where the power $\sigma$ depends on the number of transverse
directions. We are  interested in measuring the field
$\phi(X^\mu)$ for {\it real} values of its arguments $X^\mu$. It
follows that  for $\phi(X^\mu)$ to be real, the locations of the
sources $\{Y_k^0\}$ must either be real or come in complex
conjugate pairs.  In  the next section  we write down   the
prescription to obtain the closed string background associated
with an arbitrary configuration of imaginary branes, and it will
be apparent that the same reality condition is valid in full
generality.

\smallskip

For $p > 0$ there are additional on-shell open strings since the
open string tachyon can be put on-shell provided $ p^m p^m = 1$,
where $ p^m$ is the momentum along the Neumann directions. These
are   precisely the exactly marginal deformations
 studied in the original work by Callan et al. \cite{cal}.
 For example the deformation
$\lambda' \cos(X^m)$, where $X^m$ is one of the directions along
the brane,
 continuously interpolates between Neumann
and Dirichlet boundary conditions for the coordinate $X^m$. At the
critical value $\lambda' = \frac 12$, we get an array of D($p$-2)
branes localized at $X^m =2 \pi (n+\frac12)i$. The deformed
boundary state is known  for all values of $\lambda'$ \cite{cal}
and we can easily apply our prescription \refb{67} to compute $|
{ W}(\lambda') \rangle$. As $\lambda'$ varies, these states
are a family of purely closed string backgrounds. An open string
deformation, in this case $\cos(X^m)$,
 is then again re-interpreted as a closed string deformation.

\sectiono{A more general set-up}\label{768}

One important conclusion from the discussion in the previous
section  is that there is nothing {\it fundamentally} special
about the array of D-branes at $X^0 = i (n+ \frac 12) a$. Exactly
marginal open string deformations allow to move the positions of
the branes in the complex $X^0$ plane, with the only constraints
coming from the reality condition. By sending off most of the
branes to large imaginary time, we can also consider
configurations with a finite number of branes.

\smallskip

The simplest configuration consists of only one pair of branes at
$X^0 = \pm i \beta$ (with $\beta$ real, $\beta >0$). As in section
2.2, we start with the pair
in Euclidean space at $X = \pm \beta$.
The analog of \refb{sum} is
\be
\widetilde G_{pair} (X) = \frac{2 \pi}{c} e^{- \beta c} \cosh(c X) \, , \quad |X| \leq \beta\,.
\ee
Notice that the only change relative to \refb{sum} is in the
prefactor. This was foreordained since  in a finite neighborhood of
$X = 0$ both $ \widetilde G_{pair} (X)$ and $\widetilde G_{array}(X)$
 obey {\it homogeneous} wave equation $\left(\frac{d}{d X^2} -c^2 \right) \widetilde G= 0$
 and are even under $X \to - X$.  Wick rotation then gives the usual
{\it sourceless} solution 
$\sim \cos(c X^0)$.
 In Fourier transform
 \be S_{pair}(E, \dots)= \frac{\pi}{2 c}\,
e^{-\beta c} \, (\delta(E - c) + \delta(E+c))\, , \ee
 which we can write
\be
 S_{pair}(E, \dots) =  F_{pair}(E)\, {\rm Disc}_E[ \widetilde A(iE)]\,,
\ee
where
\be
 F_{pair}(E) = {\rm sign}(E) \, e^{-\beta |E|} \,.
\ee
 This result should be compared with \refb{AE}, \refb{disc},
\refb{Farray}.

\smallskip

This formula can  be  generalized  further by displacing the
D-brane pair along real time, that is at $X^0 = \alpha \pm i
\beta$. In this case  $F_{pair}(E) \to {\rm sign}(E) e^{-\beta
|E|+ i \alpha E }$. By linear superposition, an arbitrary
configuration of $M$ D-brane pairs at positions $X^0 = \alpha_k
\pm i \beta_k$,  $k=1, \dots M$, leads again to an amplitude of
the form \refb{disc}, where now the prefactor takes the general
form \be \label{su} F(E) = \sum_{k = 1}^M \, {\rm sign}(E)\,  e^{-
\beta_k |E| + i \alpha_k E }  \, \, . \ee The usual array is a
special case of this formula: With $ = \infty$, $\alpha_k = 0$,
$\beta_k = a (k- \frac 12)$ we immediately recover \refb{Farray}.

\smallskip

All the conclusions of section 3 are valid in this more general
case. The basic step is the extraction of the discontinuity in
$E$, which localizes the moduli space integration to $\rho \to 0$.
A general configuration of imaginary D-branes leads to a sphere
amplitude with an extra closed string insertion ${\cal W}$, only
the details of this insertion change.  All we have to do is to
replace $1/\sinh(\frac{|E|a}{2})$ in \refb{67} with the general
$F(E)$ given in \refb{su}. The state $| {\cal W} \rangle$ is
normalizable and has finite energy as long as all D-branes are at
distances $\beta_k > \pi$  from the real axis, generalizing the
condition $ a > 2 \pi$ that we had for the array.

\smallskip

For a generic configuration of a finite number of imaginary
D-branes, the space-time dependence of ${\cal W}$ is, however,
quite different than the case of the infinite array. This
difference is sharpest for the massless fields outside the
lightcone, $r \gg |X^0|$. For the array,
 they have the same dependence $\sim 1/r^{23-p}$ as
for a static D$p$-brane, whereas for a finite configuration they
decay one power faster $\sim 1/r^{24-p}$. This nicely dovetails
with the fact that the infinite array with $a=2\pi$ admits the
exactly marginal deformation that creates an actual D$p$- brane in
real time. The incoming and outgoing radiation that makes up
${\cal W}_{array}$ are precisely tuned to admit the deformation
that reconstructs the brane, whereas for a generic finite configuration,
brane creation would require an abrupt change of the field
asymptotics.

\smallskip
It is worth pointing out that our prescription to compute the
closed string fields associated with imaginary D-branes is
equivalent to the second-quantized point of view taken in  section
3 of \cite{mald}. They propose to obtain the wavefunction for the
closed string fields at $X^0 = 0$  by cutting open the Euclidean
path integral in the presence of D-brane sources located at
Euclidean time $\widetilde X^0  < 0$. To evaluate the expectation value of the
closed string fields in such a wave function, one needs to
construct the full configuration of sources symmetric under $\widetilde X^0
\to
- \widetilde X^0$ (obtained by simply  reflecting the sources located at
$\widetilde X^0 <
0$ to $\widetilde X^0  > 0$),
and read off the solution at $\widetilde X^0 = 0$. This is the
second-quantized version of what  we do here.

\smallskip

An interesting open question is whether as we vary
the configuration of imaginary D-branes, the state $| W \rangle$
spans the full closed string cohomology.
It is clear that to have any chance of success
we must introduce more general boundary states than the ones
considered in this paper (for example, D-branes with magnetic and
electric fields on their worldvolume), and allow
for an infinite number of imaginary branes.
 If the answer to this question is
in the affirmative then the open/closed string
duality takes a new form since at the fundamental level closed
strings and D-branes are unified.
Each closed string puncture
in a string theory amplitude could be effectively represented
by a hole in the worldsheet with appropriate boundary conditions.

\subsection{Superstring}
\label{super}

This general set-up can be generalized to  the superstring in a
straightforward way. There is now more variety of D-brane sources,
since we can consider both stable BPS D$p$-branes and unstable
non-BPS D$p$-branes. For non-BPS branes the discussion is
completely analogous to the bosonic case. One can distribute
non-BPS brane {\it pairs} freely in the complex $X^0$ plane,
subject to exactly the same reality condition as we discussed in
section 5.

\smallskip
BPS D$p$ branes  introduce on the other hand an important novelty:
They are sources of Ramond-Ramond fields. The reality condition
for the RR fields forces us to consider pairs composed of a BPS
brane at $X^0 = \alpha  +  i \beta $ and  of its {\it anti}-brane
partner at $X^0 = \alpha -i \beta$. It can be checked that the RR
fields produced by a generic configuration of such pairs are
non-vanishing only {\it inside} the lightcone, $ r \ll |X^0|$
(see also \cite{strom});
this is consistent with the fact that the configuration has zero total
RR charge,
 so we do not see long range RR fields.

\smallskip
Like in the bosonic string,
infinite `critical' arrays of branes
at
$X^0 = i \, a_{crit} \, (n + \frac12 )$
correspond to  special limits of BCFTs related to
real time processes of brane creation and annihilation.
Here $a_{crit} = \sqrt{2}\, \pi$
(this is the familiar $\sqrt{2}$ in translating
between the bosonic string and the superstring).
There are two interesting classes of examples.
One can consider a  critical array of non-BPS D($p$-1) branes
in imaginary time,
which is dual to the closed string background
related to the decay of a
 D$p$-$\bar{\rm D}p$ pair;
or a critical array of alternating BPS D($p$-1)/$\bar {\rm D}(p$-1) in imaginary time,
 related to the decay of non-BPS D$p$.
(Needless to say, given $p$,
any of these examples makes sense
in either Type IIA or IIB, but not in both).
The subject is potentially quite rich.

\sectiono{Open string field theory}

We have found an intriguing relation between D-branes in imaginary
time and purely closed backgrounds. Since D-branes admit an open
string description, this suggests that one may be able to  obtain
a  dual description of closed string theories in terms of open
strings. We have already seen in section 5 that the exactly
marginal open string deformations have a natural re-interpretation
as deformations of the closed string background. However since
there are no propagating on-shell open string degrees of freedom
on imaginary D-branes, it is clear that if such a complete
open/closed duality exists, it must involve the off-shell open
strings. In this section we offer some very brief and incomplete
speculations in this direction.

\smallskip

We would like to propose that the open string {\it field} theory
(OSFT) on a configuration of imaginary D-branes is dual to the
corresponding closed string theory. To make sense of this
speculation we must define what is the OSFT for imaginary branes.
Applying our usual strategy, we start with OSFT on standard
D-branes, and double Wick rotate. While from a first quantized
point of view it may be subtle to define the Wick-rotated open
string theory, in the second quantized approach we have the luxury
of a space-time action, which seems straightforward to
analytically continue.

\smallskip

  Let us sketch   how this may come about
in the example of the array of D(-1) branes.  We start with an
array of D(-1) branes at $\widetilde X^0 = (n + \frac 12) a$. The
open string field $\Psi_{{j} {k}}$ has Chan-Paton labels $j, k \in
{\bf Z}$ running over the positions of the D(-1) branes, and the
cubic OSFT action \cite{wit}  takes the form \be \label{OSFT}
S[\Psi] = -\frac{1}{ g_0^2}  \left( \frac{1}{2} \,  \, \sum_{jk}
\langle \, \Psi_{{jk}},   Q_B   \Psi_{{kj}} \rangle  +
\frac{1}{3} \, \sum_{jkl} \langle  \Psi_{{jk}}  ,  \Psi_{{kl}} ,
\Psi_{{lj}} \rangle  +   \sum_j   \langle \Psi_{{j}  {j}} ,  {\cal
C} \rangle \right) \,. \ee Here we have also included the
gauge-invariant open/closed vertex \cite{sha1,sha2,zwi,has,gai}
$\langle \Psi , {\cal C} \rangle$ that couples external on-shell
closed strings ${\cal C}$ to the open string field. Notice that
the string fields $\Psi_{ij}$ do not depend at all on the zero
modes of the space-time coordinates. Double Wick rotation is then
immediate.
 There is little  to do in all the purely open terms of the
 action.\footnote{Note however that one has to be careful here with
the reality conditions discussed in the previous sections
\cite{wip}.} Only  the open/closed vertex is affected. We
conjecture that the resulting action describes the non-trivial
closed string state $| W \rangle$.

\smallskip

In principle it should be possible to recover the results of
section 3 from the point of view of the second quantized Feynman
rules for the action \refb{OSFT}. In OSFT, amplitudes of external
closed string on the disk are obtained by computing expectation
values of the gauge-invariant operators discussed in
\cite{sha1,sha2,zwi,has,gai}, or in other terms by the use of the
open/closed vertex. We expect that for \refb{OSFT} such amplitudes
collapse to the region of moduli space where the open string
propagators have zero length, reducing to sphere amplitudes with
an extra insertion of $| W \rangle$. The mechanism for such a
collapse must be of a somewhat different nature than in
\cite{gai} or in \cite{nadav}, 
since here we are using the conventional BRST operator
but a highly unconventional state-space.

\smallskip
Finally, the OSFT \refb{OSFT} may provide us with new clues about
the string field theory around the tachyon vacuum. As $a \to
\infty$, the energy of $| W \rangle$
 goes to zero, and we approach the
tachyon vacuum. Interestingly, in the same limit $|W \rangle$ does
not vanish completely, but it becomes purely a zero-momentum
dilaton. Thus we recover precisely the scenario of \cite{gai}.
Since for any $a$ such an open string field
theory should make sense this may
provide us with a consistent regularization  of vacuum string field
theory,  which should be
related to the one considered in \cite{gai} by some
non-trivial field redefinition.

\sectiono{${\bf a=2\pi}$ and reconstruction of the brane  }

As discussed in section 4, for  $a=2\pi$ the normalization and the
energy associated with the closed string state $|{\cal W }\rangle$
diverge \cite{mald}.
 This singularity signals the appearance of new open
string degrees of freedom. Open strings stretched between
neighboring branes in imaginary time have conformal dimension $L_0
= (a/2\pi)^2$, which equals one for $a = 2 \pi$. The infinite
periodic array has the special property that a specific linear
combination of these marginal operators (the one which is
invariant under $X_0 \to X_0 + 2 \pi i$ and is even
under $X_0 \to - X_0$) is in fact {\it exactly} marginal\footnote{
In principle also the {\it odd} operator, `$\sinh(X^0)$', is
exactly marginal. However in the bosonic string this deformation
would bring us to the `wrong' side of the tachyon potential for
$X^0 < 0$.}. A new branch of moduli space opens up for $a = 2
\pi$. Indeed,  the theory for $a = 2 \pi$ is equivalent to the
$\lambda = \frac 12$  critical point of the BCFT \refb{iu}.
Turning on the exactly marginal open string deformation
(`$\cosh(X^0)$') we can reduce the value of $\lambda$, from
$\lambda = \frac 12$ (the purely closed string background $|W
\rangle$) to $\lambda =0$ (the usual brane with Neumann boundary
condition in time). This is the sense in which the array for  $a=
2 \pi$ is very special: it admits an exactly marginal deformation
that {\it cannot} be interpreted purely as a deformation of the
closed background.

\subsection{ Smearing and brane creation}

One can obtain many insights
 into the physics for $\lambda < 1/2$
by a simple extension of the methods in section 2. The basic idea
is that in the Euclidean BCFT with $X = - i X^0$,
 taking $\lambda < \frac12$ amounts to regulating the
delta-function sources located at $X=2\pi (n + 1/2) $ with smooth
lumps.  The precise way to make this regulation can be gleaned
from the boundary state \cite{cal} for general $\lambda$. Focusing
on the oscillator free part of the boundary state,
\be 
| {\cal B}_0 \rangle \sim \left[   1 + 2  \sum_{n=1}^\infty (-1)^n
e^{ -n \tau  }\cos( n X(0)) \right] | 0 \rangle \, , \quad \tau
\equiv - \log(\sin ( \pi \lambda)) \,. \ee By Poisson resummation,
one finds \be | {\cal B}_0 \rangle \sim \sum_{-\infty}^\infty
\tilde  j_{\tau}(X+ 2 \pi (n+ \frac12) ) | 0 \rangle \, ,\quad
\tilde  j_\tau(X) = \frac{\tau}{ \pi ( X^2 + \tau^2)}
 \,. \ee
The interpretation of this formula is clear: for $\lambda < \frac
12$, the boundary state corresponds to
 an infinite array of smeared sources $\tilde j_\tau$;
indeed $\tilde j_\tau$  is a well-known representation for
$\delta(X)$ in the limit $\tau \to 0$.

\smallskip

If we sum the source $\tilde j_\tau (X)$ over the infinite array
and then Wick rotate $X \to - i X^0$ we find \be J_\tau(X^0)
=\frac{\tanh((X^0 + \tau)/2) - \tanh((X^0 - \tau)/2)} { 4 \pi} \,.
\ee
Already at this stage we see  the crucial difference compared to
$\lambda = \frac 12$ ($\tau = 0$). Now there is   a non-zero
source
localized at {\em real} time for $|X^0| < \tau$, 
 which is to say that an unstable
D-brane  appears at $X^0=-\tau$ and disappears at $X^0=\tau$. For
later use let us record the Fourier transform of this function,
\be \label{rho} \rho_\tau(E) = \frac{ \sin( \tau E)}{\sinh(\pi E)}
\,. \ee

\smallskip

The next step is to repeat the  exercise for the {\it fields}
 generated by this smeared array. The Fourier transform of the Euclidean
source $\tilde j_\tau(X)$ is  $  e^{-\tau |P|}\, . $ Therefore, in
the notations of section 2.2, eq.\refb{i3}
 should be replaced by
\be \tilde{A}(P,...)=\frac{e^{-\tau |P|}}{P^2 + c^2} \,. \ee As in
section 2.3, we sum over the array and use the residue theorem to
write the total field as a contour integral,
 \be
\widetilde G_\tau(X) = \frac{1}{2 i} \oint_{\cal C} dP e^{iP X}
\frac{e^{-\tau |P|}}{(P^2 + c^2) \sin(\pi P)} \,. \ee If we now
move the contour over the imaginary $P$ axis as in Fig.2, and Wick
rotate, we find
\be
 \label{advret}
S_\tau(E) = \frac{1}{2 \sinh\left(\pi E\right)} \left( \frac{e^{i
\tau E}} {(E + i \epsilon)^2- c^2} -\frac{e^{-i \tau E}}{ (E - i
\epsilon)^2- c^2} \right) \, ,\ee
which we recognize as
\be S_\tau (E) = \frac{1}{2 \sinh\left(\pi E\right)} \left(
G_{ret} (E) e^{i \tau E}  - G_{adv}(E) e^{-i \tau E}  \right)\,. \
\ee
This way of writing the answer makes the space-time interpretation
manifest (see also \cite{mald}). For $X^0 < - (\tau +r)$, we have
purely {\it incoming} radiation. For $X^0 >  \tau+ r$, we have
purely {\it outgoing} radiation. Outside of these two cones (for
$(X^0)^2 - r^2 < \tau^2$), the fields are the same as  the ones
produces by a {\it static}  source. The  thickness of the
transition regions
 is of the order  of the string length. Notice that the outgoing radiation is
produced by the rapid change of the source for $X^0 \sim \tau$,
and similarly the incoming radiation is correlated  with the
change at $X^0 \sim -\tau$. This  process can be described as some
finely tuned incoming closed strings that
 create an unstable
D-brane at $X^0 = -\tau$, which  then decays at $X^0 =\tau$ into
outgoing closed strings.

\smallskip

We can also write
 \be \label{Sfey}
S_\tau(E,...)= G_F (E)\, \rho_\tau(E) + \cos(\tau E) \; \frac{
\pi}{2 c\sinh(\frac{c a}{2})} (\delta (E-c)+\delta(E+c))\, , \ee
where $G_F(E)$ is the Feynman propagator and $\rho(E)$ the Fourier
transform of the source in real time $J_\tau(X^0)$, see \refb{rho}.
 The two terms in \refb{Sfey}  have very different interpretations.
While the first term contains a propagator, and is non-zero for
any finite $E$,  the second term has support  only for $E=\pm c$.
 The second term in \refb{Sfey}
is proportional to \refb{AE}; in section 3 it was seen to
correspond to the extra insertion $| W \rangle$ in a sphere
amplitude. The first term is instead a real D-brane source,
associated to an earnest disk amplitude.

\smallskip

 In the limit
$\lambda \to \frac 12$ ($\tau \to 0$) only the second term in
\refb{Sfey} contributes, since the source $\rho_\tau(E) \to 0$. It
is a bit less obvious to see why only the first term contributes
as $\lambda  \to 0$. In this limit, $\tau \to \infty$ and since
the oscillations of both terms are controlled by the dimensionless
parameter $\tau E$, to have a non-zero contribution  $E$ must go
to zero,  hence  for any fixed finite $c$ the second term
vanishes. The fact the $E \rightarrow 0$ is consistent with the
fact  that the for $\lambda = 0$ the unstable D-brane exists for
an infinite amount of time;
 $X^0$ becomes an ordinary Neumann
direction and the total energy (with all insertions as incoming)
must vanish.

\smallskip

We see that as we vary $\lambda$, not only does the D-brane source
vary, but the closed string background (captured by the second
term in \refb{Sfey}) also changes in a very non-trivial way. This
phenomenon requires some comments.
Although the  BCFT \refb{iu} may be naively thought of as a
deformation of the open string background only, infact we need to
simultaneously change the closed string background to cancel
tadpoles. Indeed, while the operator  $\cosh(X^0)$ is  exactly
marginal in the open string sense,  one point functions of {\it
generic} on-shell closed strings have a non-trivial dependence on
$\lambda$. Of course, tadpoles for closed string operators are
also generated in the familiar case of
 time-independent boundary deformations. However
 in the time-independent case only one-point functions  of
  {\it zero-energy} on-shell bulk operators can change, whereas for a time-dependent perturbation like \refb{iu}
there are tadpoles with a non-trivial space-time dependence. As we
turn on the {cosh} deformation, the closed string background needs
to be corrected introducing  closed string matter.
 Happily, as \refb{Sfey} shows, this seems to be
automatically taken care of by
the analytic continuation procedure.

\smallskip

The discussion in this section  has been at the level of the
discussion of section 2. It would be very interesting to repeat
the analysis of section 3 and compute scattering amplitudes of
closed strings for general $\lambda$. It would also be of interest
to find the behavior of the open string spectrum for $0<\lambda
<\frac 12$.

\sectiono{From Choptuik to Gregory-Laflamme}

As was explained in section \ref{aa}  the massless sector of $|
{W} \rangle$ is a spherically symmetric dilaton wave in $26-p$
dimensions. In the limit that the Newton constant $G_N$ goes to
zero  the linearized solution is an exact solution to the
gravity-dilaton  equations of motion. As the coupling constant is
turned on the non-linearity of the equation of motion becomes more
and more important. With spherical symmetry much is known about
the non-linear aspects of the system. In particular,  
Choptuik showed   \cite{cho}, via numerical analysis, that a universal behavior
occurs at the critical point where black hole formation first
occurs. A crude  summary of his results is the following. Consider
a spherically symmetric wave of a massless scalar field
\be\label{bhn} \eta \, \phi(r,X^0). \ee
The strength of the non-linear effects of the gravitational
back-reaction of the wave grows with the overall
coefficient $\eta$. When $\eta\rightarrow
0$ the linearized approximation is exact while for
$\eta\rightarrow\infty$ a large black hole will be formed, with an
exponentially small amount of energy escaping the black hole
formation as an outgoing radiation.\footnote{The system is
classical so this radiation is classical and  should not be
confused with  Hawking radiation.} Therefore, for a given field
profile
$\phi(r, X^0)$ there is a special value $\eta^*$ where the black hole
formation first takes place. While $\eta^*$ certainly depends on
the details of $\phi$,
the time evolution of the system for $\eta\rightarrow
\eta^*$ admits scaling behavior that is fixed by a certain
constant, $\Delta$.
This constant is universal in the sense that
it does not depend on $\phi$, but it can depend for example
on the number of spatial directions. Another critical
exponent $\delta$ is related
to the scaling of the mass of the black hole near
the transition. These fascinating results have been
explored quite extensively in the last decade.

\smallskip

In this section we take advantage of the fact that
a spherically symmetric dilaton wave can be viewed as a configuration
of D-branes located in imaginary time to relate  Choptuik's findings to a
phase transition somewhat reminiscent of the Gregory-Laflamme
instability \cite{Greg}. For simplicity we will mostly phrase
the discussion in the context of the bosonic string,
ignoring as usual the closed string tachyon.
The extension to the superstring of the
scaling arguments given below is straightforward.

\smallskip

The first step
is to understand how
 to get from our $| {W} \rangle$ a background containing only
a classical dilaton wave.
It is clear that to suppress
massive closed string modes we need to
take $\frac{a}{l_s}\rightarrow \infty$. (Here we have restored
the string length $l_s$, that was set to one in the rest of the paper).
However the total energy of the
wave will then go to zero
as $1/a^{25-p}$  in this limit (see \refb{barE}),
and a  black hole will not be formed this way.
A simple way to obtain a configuration
with enough energy without exciting the
massive modes is to increase the number of branes at each point
$X^0=ia(n+\frac12 )$,  as in the discussion at
 the end of section \ref{aa}.
With $N$ D-brane at each point the  particles density, $\bar n
/V$, and the  energy density $ \epsilon =\bar E /V$ ($V$ is the
volume along the wave) of the wave for large $a$ are (from
\refb{barE})
\be \frac{\bar n}{V} \sim \frac{N^2
l_s^{24-2p}}{a^{24-p}},\;\;\;\;\; \epsilon \sim \frac{N^2
l_s^{24-2p}}{a^{25-p}}\,.\ee
Let us now state the exact conditions that must be satisfied in
the limit we want to take:

\begin{enumerate}

\item
The string coupling $g_s$ should go to zero
to suppress quantum effects, and the
total number of particles $\bar n$ should go to
infinity so that $|{ W}\rangle $ is well described by a classical wave.

\item $\frac{l_s}{a} \to 0 $ to ensure that
the massive closed strings decouple.

\item The gravitational radius associated with the total energy of
the wave, \be r_G=(G_N  \epsilon)^{\frac{1}{23-p}},\ee 
 (here $G_N$ is the
Newton constant in 26 dimensions)  should be comparable to the wavelength $a$.
If we define the dimensionless parameter
\be\label{212}\zeta =\frac{r_G}{a}=\frac{
(\,g_s^2 N^2
l_s^{48-2p}/a^{25-p}\,)^{\frac{1}{23-p}} }{a},\ee
then in our set-up
$\zeta $ plays the role of $\eta$  in \refb{bhn}
That is, for small $\zeta$ the linearized approximation is valid
and a black hole will not be formed while for large $\zeta$
non-linear effects are important and lead to black hole
formation. So the critical value $\zeta^*$ is a number of order one.

\end{enumerate}
It is easy to verify that these conditions are satisfied in the
following limit
\be\label{top} a\;\; \mbox{fixed},\;\;\;\;\; l_s\rightarrow
0,\;\;\;\;\;g_s\propto l_s^{\beta},\;\;\;\;\;\; N\propto
1/l_s^{24+\beta-p},\;\;\;\;\;\;\beta>0.\ee
-So far what we have done is to take a certain `decoupling' limit
that leaves  us with the classical picture of black-hole
formation. The obvious question to ask is what does
this mean  from the D-brane point of view. In particular, does anything
special happen at $\zeta = \zeta^*$ in the array of D($p$-1) branes
located at $\widetilde X^0 = (n + \frac 12) a$, {\it i.e.}
in the spatial array before the double Wick rotation?

\smallskip

The gravitational radius associated with N D-branes located at
each site is
\be\label{ds} l_G\sim (g_s N)^{\frac{1}{24-p}} \, l_s.\ee
When $l_G$ is much smaller than the distance between the
D-branes, $a$, the gravitational interactions between them is
small and they can be considered as separated points. However,
when $l_G $ is larger than $a$ the gravitational interaction
between them is so strong that effectively they form a
black hole along $\widetilde X^0$.
It is easy to see from (\ref{ds},\ref{212})
that this transition occurs at the same point
as the wave to black hole transition.
Since the two processes are
related by Wick rotation it is very tempting to suspect that there
is a precise numerical relation  between the exponents of
\cite{cho} and the exponents, yet to be found, in the
phase transition just described.

\smallskip

It should be stressed that although
the transition considered here is somewhat reminiscent
of the Gregory-Laflamme
instability, there are also
obvious differences. In the Gregory-Laflamme
case
one starts with a black $p$-brane and finds,
at the level of linearized equations of motion,
an instable mode; condensation of this tachyon is then conjectured
to lead to an array of black ($p$-1)-branes.
(This expectation
however may even be false,
see {\it e.g.} \cite{hor,gub,kol} for recent discussions).
In our case we start from
the array and we turn on
the coupling constant to eventually form a (possibly non-uniform)
black $p$ brane, so
we are going in the opposite direction.
More crucially,
we do not start with an array of {\it black} $p$-branes. The D-branes are
extremal to begin with, in the sense that they
have no finite area horizon. In
the process of turning on the coupling constant the radius of
their zero area horizon grows until they meet. At that point a
combined finite area horizon will be formed.

\smallskip

Let us
illustrate this in the context of the superstring. In this case there
are no closed string tachyons so the whole discussion makes more sense.
We can consider either an array of BPS D($p$-1)-branes,
which must then be alternating as D/$\bar {\rm D}$,
or an array of non-BPS D$(p$-1) branes.
 Consider for example
an array of D3/$\bar {\rm  D}$3 in Type IIB. For small $\zeta$ the
D3 branes interact weakly and so they can be treated as separated
stable objects. As we turn on $\zeta$ their gravitational radius
start to touch. At that stage the fact that we have alternating
branes anti-branes is crucial since the charges can annihilated to
form a black 4-brane. This black 4-brane does not carry any 6-form
charge (we are in type IIB).

\smallskip

It would be extremely interesting to understand Choptuik's
critical behavior from the point of view of the dual open string field
theory discussed very briefly in section 7. Very
schematically, the way this could work is as follows.
From the open string field theory point of view the classical
non-linear gravity dynamics is obtained in the loop expansion.
We expect that the lightest open
string mode stretched between the branes, which
 at the classical level has large positive $m^2$,
becomes {\it tachyonic} for
 $\zeta > \zeta^*$ due to quantum open string effects.
It may
be possible to see this effect  in a one-loop computation,
in the spirit of \cite{col}. If this scenario is
correct, the universal physics at the transition point
would be completely captured by a massless field,
and we would have an explanation
of gravitational critical behavior in terms
of a second order phase transition in
the dual OSFT.

\bigskip

\bigskip

\bigskip

\noindent {\bf Acknowledgements}

We thank C. Callan, A. Hashimoto, I. Klebanov, H. Liu,
J. Maldacena, J. McGreevy, H. Ooguri, A. Sen, C. Vafa
and H. Verlinde for discussions.
This material is based upon work supported by the National Science
Foundation under Grant No. PHY 9802484. Any opinions, findings,
and conclusions or recommendations expressed in this material are
those of the author and do not necessarily reflect the views of
the National Science Foundation.



\begin{thebibliography}{99}

\bibitem{gut}
M.~Gutperle and A.~Strominger, ``Spacelike branes,'' JHEP {\bf
0204}, 018 (2002) [arXiv:hep-th/0202210].

\bibitem{cosh}
A.~Sen,
``Rolling tachyon,''
JHEP {\bf 0204}, 048 (2002)
[arXiv:hep-th/0203211].




\bibitem{s}
A.~Sen, ``Tachyon matter,'' JHEP {\bf 0207}, 065 (2002)
[arXiv:hep-th/0203265].


\bibitem{4}
A.~Sen, ``Field theory of tachyon matter,'' Mod.\ Phys.\ Lett.\ A
{\bf 17}, 1797 (2002) [arXiv:hep-th/0204143].



\bibitem{3}
A.~Sen, ``Time evolution in open string theory,'' JHEP {\bf 0210},
003 (2002) [arXiv:hep-th/0207105].


\bibitem{2}
P.~Mukhopadhyay and A.~Sen, ``Decay of unstable D-branes with
electric field,'' JHEP {\bf 0211}, 047 (2002)
[arXiv:hep-th/0208142].


\bibitem{8a}
A.~Strominger, ``Open string creation by S-branes,''
arXiv:hep-th/0209090.




\bibitem{1}
A.~Sen, ``Time and tachyon,'' arXiv:hep-th/0209122.



\bibitem{8}
F.~Larsen, A.~Naqvi and S.~Terashima, ``Rolling tachyons and
decaying branes,'' JHEP {\bf 0302}, 039 (2003)
[arXiv:hep-th/0212248].


\bibitem{7}
M.~Gutperle and A.~Strominger, ``Timelike boundary Liouville
theory,'' arXiv:hep-th/0301038.


\bibitem{strom}
A.~Maloney, A.~Strominger and X.~Yin,
``S-brane thermodynamics,''
arXiv:hep-th/0302146.


\bibitem{mald}
N.~Lambert, H.~Liu and J.~Maldacena,
``Closed strings from decaying D-branes,''
arXiv:hep-th/0303139.


\bibitem{9}
B.~Chen, M.~Li and F.~L.~Lin, ``Gravitational radiation of rolling
tachyon,'' JHEP {\bf 0211}, 050 (2002) [arXiv:hep-th/0209222].

\bibitem{10}
S.~J.~Rey and S.~Sugimoto, ``Rolling tachyon with electric and
magnetic fields: T-duality approach,'' arXiv:hep-th/0301049.

\bibitem{11}
N.~Moeller and B.~Zwiebach, ``Dynamics with infinitely many time
derivatives and rolling tachyons,'' JHEP {\bf 0210}, 034 (2002)
[arXiv:hep-th/0207107].

\bibitem{12}
S.~Sugimoto and S.~Terashima, ``Tachyon matter in boundary string
field theory,'' JHEP {\bf 0207}, 025 (2002)
[arXiv:hep-th/0205085].

\bibitem{13}
J.~A.~Minahan, ``Rolling the tachyon in super BSFT,'' JHEP {\bf
0207}, 030 (2002) [arXiv:hep-th/0205098].

\bibitem{14}
T.~Okuda and S.~Sugimoto,
``Coupling of rolling tachyon to closed strings,''
Nucl.\ Phys.\ B {\bf 647}, 101 (2002) [arXiv:hep-th/0208196].


\bibitem{15}
J.~Kluson, ``Exact solutions in open bosonic string field theory
and marginal  deformation in CFT,'' arXiv:hep-th/0209255.

\bibitem{16}
I.~Y.~Aref'eva, L.~V.~Joukovskaya and A.~S.~Koshelev, ``Time
evolution in superstring field theory on non-BPS brane. I: Rolling
tachyon and energy-momentum conservation,'' arXiv:hep-th/0301137.

\bibitem{e}
A.~Ishida and S.~Uehara, ``Rolling down to D-brane and tachyon
matter,'' JHEP {\bf 0302}, 050 (2003) [arXiv:hep-th/0301179].

\bibitem{shata}
S.~L.~Shatashvili,
``On field theory of open strings, tachyon condensation and closed  strings,''
arXiv:hep-th/0105076.


\bibitem{gai}
D.~Gaiotto, L.~Rastelli, A.~Sen and B.~Zwiebach, ``Ghost structure
and closed strings in vacuum string field theory,''
arXiv:hep-th/0111129.


\bibitem{vsft}L.~Rastelli, A.~Sen and B.~Zwiebach,
``String field theory around the tachyon vacuum,''
Adv.\ Theor.\ Math.\ Phys.\  {\bf 5}, 353 (2002)
[arXiv:hep-th/0012251].


\bibitem{cal} C.~G.~Callan, I.~R.~Klebanov,
A.~W.~Ludwig and J.~M.~Maldacena, ``Exact solution of a boundary
conformal field theory,'' Nucl.\ Phys.\ B {\bf 422}, 417 (1994)
[arXiv:hep-th/9402113].

\bibitem{pol}
J.~Polchinski and L.~Thorlacius,
``Free Fermion Representation Of A Boundary Conformal Field Theory,''
Phys.\ Rev.\ D {\bf 50}, 622 (1994)
[arXiv:hep-th/9404008].

\bibitem{cho}
M.~W.~Choptuik, ``Universality And Scaling In Gravitational
Collapse Of A Massless Scalar Field,'' Phys.\ Rev.\ Lett.\  {\bf
70}, 9 (1993).

\bibitem{Greg}
R.~Gregory and R.~Laflamme, ``Black Strings And P-Branes Are
Unstable,'' Phys.\ Rev.\ Lett.\  {\bf 70}, 2837 (1993)
[arXiv:hep-th/9301052].





\bibitem{kt}
I.~R.~Klebanov and L.~Thorlacius, ``The Size of p-Branes,'' Phys.\
Lett.\ B {\bf 371}, 51 (1996) [arXiv:hep-th/9510200].

\bibitem{hk}
A.~Hashimoto and I.~R.~Klebanov, ``Decay of Excited D-branes,''
Phys.\ Lett.\ B {\bf 381}, 437 (1996) [arXiv:hep-th/9604065].

\bibitem{polchinski} J. Polchinski, "String Theory Vol. 1"
Cambridge University Press (1995).

\bibitem{zwi2}
B.~Zwiebach, ``Closed string field theory: Quantum action and the
B-V master equation,'' Nucl.\ Phys.\ B {\bf 390}, 33 (1993)
[arXiv:hep-th/9206084].

\bibitem{068}
P.~Di Vecchia, M.~Frau, I.~Pesando, S.~Sciuto, A.~Lerda and
R.~Russo, ``Classical p-branes from boundary state,'' Nucl.\
Phys.\ B {\bf 507}, 259 (1997) [arXiv:hep-th/9707068].

\bibitem{gsw} M. Green, J. Schwarz and E.Witten, "Superstring
Theory Vol. 1" Cambridge University Press (1987).

\bibitem{wit}
E.~Witten, ``Noncommutative Geometry And String Field Theory,''
Nucl.\ Phys.\ B {\bf 268}, 253 (1986).


\bibitem{sha1}
J.~A.~Shapiro and C.~B.~Thorn, ``Brst Invariant Transitions
Between Closed And Open Strings,'' Phys.\ Rev.\ D {\bf 36}, 432
(1987).

\bibitem{sha2}
J.~A.~Shapiro and C.~B.~Thorn, ``Closed String - Open String
Transitions And Witten's String Field Theory,'' Phys.\ Lett.\ B
{\bf 194}, 43 (1987).


\bibitem{zwi}
B.~Zwiebach, ``Interpolating string field theories,'' Mod.\ Phys.\
Lett.\ A {\bf 7}, 1079 (1992) [arXiv:hep-th/9202015].



\bibitem{has}
A.~Hashimoto and N.~Itzhaki, ``Observables of string field
theory,'' JHEP {\bf 0201}, 028 (2002) [arXiv:hep-th/0111092].



\bibitem{wip} Work in progress.


\bibitem{nadav}
N.~Drukker,
``Closed string amplitudes from gauge fixed string field theory,''
arXiv:hep-th/0207266.


\bibitem{hor}
G.~T.~Horowitz and K.~Maeda, ``Fate of the black string
instability,'' Phys.\ Rev.\ Lett.\  {\bf 87}, 131301 (2001)
[arXiv:hep-th/0105111].

\bibitem{gub}
S.~S.~Gubser, ``On non-uniform black branes,'' Class.\ Quant.\
Grav.\  {\bf 19}, 4825 (2002) [arXiv:hep-th/0110193].

\bibitem{kol}
B.~Kol, ``Topology change in general relativity and the black-hole
black-string  transition,'' arXiv:hep-th/0206220.

\bibitem{col}
S.~R.~Coleman and E.~Weinberg, ``Radiative Corrections As The
Origin Of Spontaneous Symmetry Breaking,'' Phys.\ Rev.\ D {\bf 7},
1888 (1973).


\end{thebibliography}
\end{document}